\documentclass{article}
\usepackage[OT1]{fontenc}
\usepackage[latin9]{inputenc}
\usepackage{array}
\usepackage{varioref}
\usepackage{float}
\usepackage{booktabs}
\usepackage{units}
\usepackage{url}
\usepackage{amsmath}
\usepackage{amsthm}
\usepackage{amssymb}
\usepackage{graphicx}
\usepackage{setspace}
\usepackage[authoryear]{natbib}
\usepackage[unicode=true]
 {hyperref}

\makeatletter

\DeclareTextSymbolDefault{\textquotedbl}{T1}
\providecommand{\tabularnewline}{\\}

\numberwithin{equation}{section}
\numberwithin{figure}{section}

\usepackage{arxiv}
\usepackage{amsmath}
\usepackage{authblk}
\usepackage{layout}
\usepackage{enumitem}
\setlist{nolistsep}

\providecommand{\keywords}[1]{\textbf{\textit{Keywords:}} #1}

\@ifundefined{showcaptionsetup}{}{%
 \PassOptionsToPackage{caption=false}{subfig}}
\usepackage{subfig}
\makeatother

\begin{document}
\markboth{N.G.P. Den Teuling et al.}{Methods for longitudinal clustering}
\title{Clustering of longitudinal data: A tutorial on a variety of approaches}
\author[1,3]{N.G.P. Den Teuling}
\author[2,3]{S.C. Pauws}
\author[1]{E.R. van den Heuvel}
\affil[1]{Dep.\ of Mathematics and Computer Science, Eindhoven University of Technology, Eindhoven, the Netherlands}
\affil[2]{Department Communication and Cognition, Tilburg University, Tilburg, the Netherlands}
\affil[3]{Philips Research, Eindhoven, the Netherlands}
\maketitle
\begin{abstract}
During the past two decades, methods for identifying groups with different
trends in longitudinal data have become of increasing interest across
many areas of research. To support researchers, we summarize the guidance
from the literature regarding longitudinal clustering. Moreover, we
present a selection of methods for longitudinal clustering, including
group-based trajectory modeling (GBTM), growth mixture modeling (GMM),
and longitudinal $k$-means (KML). The methods are introduced at a
basic level, and strengths, limitations, and model extensions are
listed. Following the recent developments in data collection, attention
is given to the applicability of these methods to intensive longitudinal
data (ILD). We demonstrate the application of the methods on a synthetic
dataset using packages available in R.

\end{abstract}

\section{Introduction}

The analysis of longitudinal data is prominent in correlational studies
that look for correspondence between observations of the same variables
over extended period of time, such as substance use or mental health
in psychology, recidivism behavior in sociology, and relapse or medication
adherence in medicine. Longitudinal studies enable researchers to
assess and study changes over time of the variables of interest. With
the increasing capabilities of data collection and storage, more and
more longitudinal studies are designed to involve a large number of
repeated measurements of the same variable per subject over time.
When a considerable number of observations are available, the data
is commonly referred to as intensive longitudinal data (ILD) \citep{walls2006models}.
ILD has the advantage of allowing for a more granular assessment of
change over time, especially at the subject level.

Analyzing longitudinal data requires models that take the structure
of the data into account. The assessment of variability is key, as
no two subjects are identical. In addition to the presence of measurement
variability within each subject, models should account for differences
(i.e., heterogeneity) between subjects. For example, in the analysis
of medication adherence, subjects may exhibit considerably different
levels of adherence over time. An example of such a modeling approach
is multilevel modeling. Here, the model describes the mean trend (i.e.,
longitudinal pattern), and captures the differences between subjects
by modeling the subject-specific deviations from the trend.

In studies with considerable between-subject variability or non-normal
deviations from the trend, subjects may exhibit large deviations,
to the point that the mean trend may not be representative of the
longitudinal patterns of the subjects \citep{hamaker2012researchers}.
An intuitive alternative approach is to represent the differences
across subjects in terms of a set of common trends. This way, the
subject-specific deviations are reduced to the nearest trend. This
approach is generally referred to as longitudinal clustering, and
involves the automatic discovery of groups of subjects with similar
longitudinal characteristics. Longitudinal clustering is of interest,
for example, in behavioral studies, where subjects can exhibit a range
of behaviors that are due to various unobserved factors, resulting
in structural deviations. We shall use the level of adherence of sleep
apnea patients to positive airway pressure (PAP) therapy as the running
example in this work. Factors such as perceived importance, self-efficacy,
personality traits, claustrophobia, and many more have been shown
to affect the level of adherence to PAP therapy \citep{cayanan2019review},
resulting in a spectrum of longitudinal patterns across patients.

In this tutorial, we present a review of the literature on methods
for clustering longitudinal data. While there are several aspects
to modeling longitudinal data, we focus on the discovery of subgroups
with different forms of longitudinal variations. Moreover, we summarize
the guidance from literature on how to conduct such a longitudinal
cluster analysis. Several types of methods have been proposed over
the past two decades for clustering longitudinal data. Our intent
is to assist the reader in making an informed decision on which method
to apply in their cluster analysis, and to acquaint the reader with
the available methodologies for longitudinal clustering. We describe
each method along with its assumptions, advantages, and practical
limitations. Secondly, we cover the topic of model specification,
with a focus on the number of clusters needed to best represent the
data. We survey the commonly used metrics and approaches to identify
the most appropriate number of subgroups. Lastly, the methods are
demonstrated on a synthetic dataset inspired by a real-world example
in the context of daily PAP therapy adherence of patients with sleep
apnea \citep{aloia2008time}. We will use this dataset to highlight
differences in the assumptions of the methods and the estimation,
as well as to show how to specify and apply each method.

The selection of methods has been based on prevalence and with the
aim of creating a varied selection with different strengths and limitations.
The variety of methods enables readers to select the most appropriate
method for their case study. Moreover, we only considered methods
that are applicable for identifying univariate longitudinal patterns
of change, and have a publicly available implementation in R. Relevant
papers were identified via keyword searches and the snowball method.
While we present the methods for the purpose of analyzing ILD, each
of the methods are applicable to repeated measures data to some degree.
The application of longitudinal clustering is becoming more commonplace.
Based on a conservative keyword search\footnote{A systematic search was performed per decade using Web of Science.
Articles must contain the keyword ``longitudinal'', and one of the
keywords ``mixture'', ``latent-class'', ``clustering'', or ``group-based''.}, we observe a considerable increase in the number of publications
concerning longitudinal clustering in different fields over time,
from 37 publications in the nineties, to 273 publications between
2000\textendash 2009, and 1,257 publications between 2010\textendash 2019.

\paragraph*{Terminology}

As the scope of this review is intended to be interdisciplinary,
we describe the key terms used in this paper, and list the commonly
used alternative terms. We explain the topic of longitudinal cluster
analysis in the context of clustering subjects over time, but the
methodology applies equally well to any application involving repeated
measures data, e.g. modeling devices, animal growth, or accident rates.

At the subject level, the sequence of longitudinal observations are
commonly referred to as a trajectory, a time series, a temporal pattern,
a curve, a trend, or a dynamic. Due to the frequency of measurement
in the case of ILD, measurements are not necessarily equidistant in
time. Moreover, subjects can have different non-corresponding times
of measurement, and the number of measurements may vary. With this
in mind, we define the trajectory of a subject $i$ as a sequence
of $n_{i}$ observations by

\begin{equation}
\boldsymbol{y}_{i}=\{y_{i,1},y_{i,2},...,y_{i,n_{i}}\},
\end{equation}
where the observation $y_{i,j}$ is taken at time $t_{i,j}\in\mathbb{R}$.

By clustering, we refer to the definition of a cluster analysis from
the field of machine learning, specifically that of unsupervised learning,
where data is grouped (i.e., clustered) based on similarity, and the
group definitions and assignments are not known in advance. In the
field of statistics a distinction is made between known clusters and
unobserved clusters. In the former case, subjects are stratified based
on a known nominal factor, for example, by assigning subjects to subgroups
based on age or sex. Unknown clusters are commonly referred to as
latent (i.e., hidden) classes, groups, profiles, or clusters. In this
paper, we use the term cluster as referring to the unobserved type.

Longitudinal clustering can be regarded as a specific area of time
series clustering that is specifically concerned with the identification
of common patterns of change or state changes throughout a longitudinal
study. Whereas the scope of time series clustering extends to the
modeling and assessment of any temporal similarity for any type of
time series data \citep{aghabozorgi2015time}. Moreover, it includes
the identification of temporal subsequences within time series.

\paragraph*{Overview}

The paper is organized as follows. We begin by elaborating on the
case study in Section \ref{sec:Case-study}. In Section \ref{sec:Background},
we first summarize the concept of multilevel modeling as a precursor
to modeling subgroups. Furthermore, we explain the concept of clustering;
both philosophically and practically. The selected methods are described
and demonstrated in Section \ref{sec:Methods}. We outline the recommended
steps involving a longitudinal cluster analysis in Section \ref{sec:guide}.
Lastly, Section \ref{sec:Discussion} discusses the findings from
the case study in addition to the general challenges, limitations,
and future work of longitudinal clustering.

\section{Case study\label{sec:Case-study}}

We use the case study in order to illustrate the longitudinal cluster
methods, and to contrast the strengths and limitations of the methods
in a practical setting. The longitudinal methods are applied to a
synthetic dataset, which facilitates a more detailed comparison between
methods, and enables a fully reproducible and transparent demonstration. The
data is generated from the population characteristics and groups as
reported by \citet{aloia2008time}, who investigated patterns of daily
time on therapy among 71 obstructive sleep apnea patients in their
first year of therapy. The synthetic dataset and analysis code are
provided in the supplementary materials.

Sleep apnea is a common chronic disorder. Patients suffer from frequent
paused or diminished breathing during their sleep, resulting in fragmented
sleep and overall poor sleep quality. Sleep apnea is commonly treated
using positive-airway pressure (PAP) therapy. This involves a device
that assists the patient in breathing during sleep by supplying positive
air pressure through a mask worn by the patient. Patients are required
to use the device every time they sleep. Considering the inconveniences
and difficulties patients can face with the therapy, some patients
struggle to comply with the therapy for longer periods of time, whereas
others do well. The progression of the therapy is determined by many
factors, e.g. the initial perception patients have of the therapy,
the coping ability of the patient, and social support \citep{weaver2008adherence,cayanan2019review}.
An effective treatment can only be ensured when patients are compliant
to the therapy, where the threshold for therapy compliance is usually
set at 4 hours of therapy per day, but PAP use for longer than 6 hours
has been shown to have positive effects \citep{weaver2008adherence}.
The patterns of change in usage hours are therefore of interest. Most
past studies have treated the patient population as being homogeneous,
whereas others have attempted to stratify the population in order
to address the differences in therapy adherence over time between
subjects \citep{aloia2008time,babbin2015identifying}.

\citet{aloia2008time} modeled the trajectories of daily hours on
therapy of each patient in terms of an intercept, slope, variance,
autocorrelation, and number of attempted days. Seven clusters were
manually identified using two expert raters. The cluster of Good users
(24\%) have a high number of therapy days and a high average hours
of usage (6.6 hours). Slow improvers (13\%) have an initially low
number of hours early in therapy but increased over time, whereas
the Slow decliners (14\%) exhibit the opposite pattern. Variable users
(17\%) have a lower average usage (5 hours), and showed fluctuations
in adherence over time. Occasional attempters (8\%) have low attempt
probability and low hours of use (3.2 hours), but the patients did
continue therapy. Lastly, a sizable proportion of patients prematurely
stopped with therapy, as represented by the Early drop-outs (13\%)
and Non-users (11\%).

\begin{figure}
\caption{\label{fig:syndata}Generated dataset with 500 patients, including
non-attempted days as zero hours.}

\subfloat[The mean cluster trajectories.]{\includegraphics{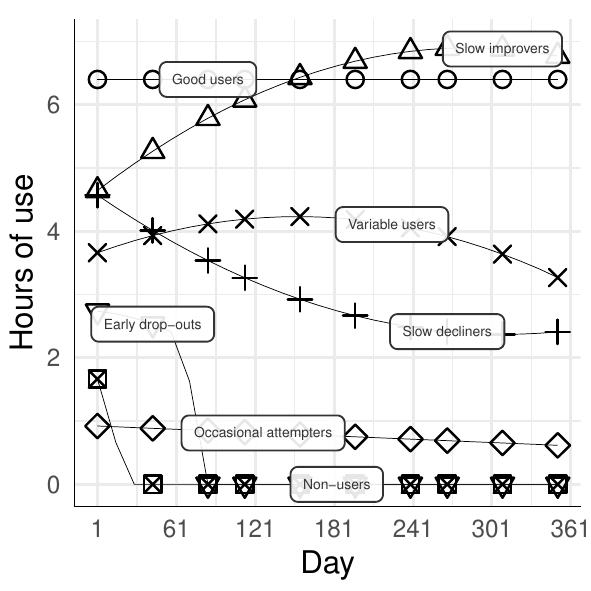}}\hfill{}\subfloat[The 14-day averaged trajectories per cluster.]{

\includegraphics{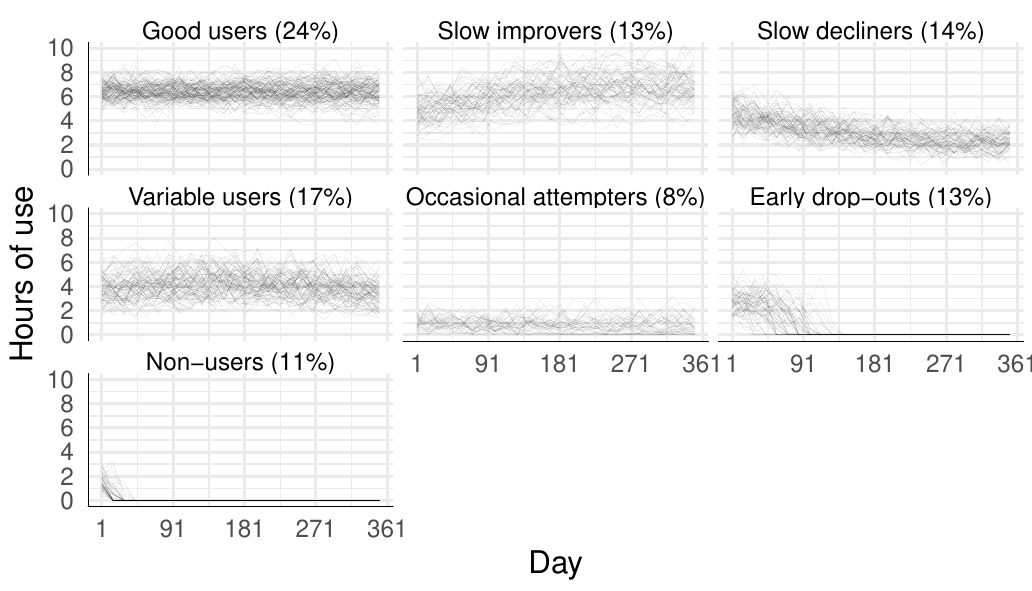}}
\end{figure}

We utilize the reported patient and cluster statistics to generate
500 patient trajectories, with each patient comprising at most 361
observations. The trajectories are generated according to the original
cluster proportions, and each trajectory is assigned a random deviation
in intercept and slope from its respective cluster. Considering the
scope on identifying patterns of change, we introduce a second-order
term in the cluster trajectory shapes to evaluate whether the methods
are able to recover these shapes. The cluster coefficients used to
generate the trajectories are reported in Table \ref{tab:syndata}.
Due to the considerable computation time of the mixture methods, we
downsampled the generated data to a biweekly average, resulting in
26 observations per patient, with 13,000 observations in total. The
cluster trajectories and downsampled individual trajectories are visualized
in Figure \ref{fig:syndata}. Overall, 21\% of biweekly observations
are zero, and the mean non-zero usage is 4.6 hours ($\sigma=2.1$
hours).

\subsection{Evaluation}

All methods are evaluated in R 3.6.3 using freely available packages
\citep{rcoreteam2020r}. Each method is evaluated for 1 to 8 clusters
in order to assess the number of clusters that correspond to the most
representative solution of each method. If available for the respective
method, we use the Bayesian information criterion (BIC) in order to
guide the identification of the most appropriate number of clusters
per method. It is defined by $\mathrm{BIC}=p\log n-2\log\hat{L}$,
where $\hat{L}$ denotes the likelihood of the candidate model, $p$
is the number of model parameters, and $n$ are the number of observations.
The BIC is one of the most widely used metrics in longitudinal clustering
\citep{van2017grolts}. A lower BIC indicates a more representative
model for the data. The BIC includes a penalty factor for model complexity,
resulting in a higher score for a model with more parameters. If the
BIC values are similar between adjacent solutions, we base the final
choice for the number of clusters on a subjective analysis of the
variety in patterns identified by the methods \citep{nagin2010groupclinical}.

In case the BIC is not available for the respective longitudinal cluster
method (i.e., there is no model likelihood), we apply the average
silhouette width (ASW) \citep{rousseeuw1987silhouettes}. The ASW
is a data-based measure of class separation. The silhouette value
measures the similarity of an object to the objects in its assigned
cluster, relative to the similarity of the other clusters. It is expressed
as a score between -1 and 1, where a higher value indicates a greater
similarity to the assigned cluster. The ASW is obtained by averaging
the silhouette values of all trajectories. The topic of selecting
the number of clusters is discussed in more detail in Section \ref{sec:numgroups}.
\begin{table}
\caption{\label{tab:syndata}Group coefficients for generating the trajectories.
Values enclosed in parentheses denote the standard deviation of the
random effects. The attempt probability is conditional on the patients
still being on therapy. The early drop-outs and non-users are modeled
to stop prematurely at day 80 (30) and day 20 (10), respectively.}

\noindent \centering{}%
\begin{tabular}{rcr@{\extracolsep{0pt}.}lr@{\extracolsep{0pt}.}lr@{\extracolsep{0pt}.}lr@{\extracolsep{0pt}.}lc}
{\footnotesize{}Cluster} & $\pi$ & \multicolumn{2}{c}{$\beta_{0}$} & \multicolumn{2}{c}{$\beta_{1}\times10^{2}$} & \multicolumn{2}{c}{$\beta_{2}\times10^{4}$} & \multicolumn{2}{c}{$\sigma^{2}$} & $p_{\mathrm{attempt}}$\tabularnewline
\midrule
{\footnotesize{}Good users} & 24\% & 6&6 (.54) & 0&0 (.16) & 0&0 & 2&0 (.82) & 97\%\tabularnewline
{\footnotesize{}Slow improvers} & 13\% & 4&8 (1.0) & 1&7 (.16) & -0&30 & 3&6 (1.3) & 94\%\tabularnewline
{\footnotesize{}Slow decliners} & 14\% & 6&1 (.63) & -1&9 (.14) & 0&30 & 3&2 (.85) & 77\%\tabularnewline
{\footnotesize{}Variable users} & 17\% & 4&4 (.87) & 0&96 (0.0) & -0&30 & 3&4 (1.2) & 82\%\tabularnewline
{\footnotesize{}Occasional attempters} & 8\% & 3&2 (1.1) & -0&30 (.91) & 0&0 & 3&6 (1.8) & 29\%\tabularnewline
{\footnotesize{}Early drop-outs} & 13\% & 4&0 (1.1) & -0&14 (1.0) & -1&0 & 5&0 (2.6) & 69\%\tabularnewline
{\footnotesize{}Non-users} & 11\% & 2&5 (.93) & -1&5 (1.0) & -1&0 & 3&0 (1.7) & 70\%\tabularnewline
\bottomrule
\end{tabular}
\end{table}

\section{\label{sec:Background}Background}

Prior to performing an exploratory cluster analysis, it is worthwhile
to consider the case where the data comprises no clusters, i.e., the
case where there is only a single cluster \citep{greenberg2016criminal,bauer2007observations,sher2011alcohol}.
We therefore begin by describing regression modeling, and how regression
models can capture heterogeneity without the need for clusters.

Subjects can be considered as independent sources of variation, with
each subject having, for example, their own mean response level, change
over time, or measurement variability. Under this assumption, and
given that subjects have a sufficient number of observations (as is
typically the case with ILD), subjects can be modeled independently
using established methods from the field of time series analysis \citep{liu2017person}.
This is referred to as a two-step or bottom-up approach. The individual
time series are commonly represented using linear regression or autoregressive
models.

Aside from between-subject variability, there may be other sources
of variation in the data. One can think of the measurement device
used by the subject having a certain measurement error, which is shared
across subjects using the same device. Another common source of variability
are the different sites at which measurements are collected. Mixed
modeling \citep{hartley1967maximum,laird1978nonparametric} enables
researchers to assess subject-specific effects, and to decompose the
variability in the data. It is also commonly referred to as hierarchical
modeling, random effects modeling, random coefficient modeling, and
variance component modeling. The model describes the population-level
effects, referred to as fixed effects, and describes a part of the
subject-specific observations in terms of a structural deviation from
the fixed effects. The subject-specific deviations are a source of
variation as the deviations cannot be fully explained in terms of
covariates, and therefore are treated as random variables, also referred
to as a random effects or latent variables.

\begin{figure}
\noindent \begin{centering}
\caption{A linear mixed model derived from 12 trajectories with random intercept
$\sim N(3,5)$, random slope $\sim N(\nicefrac{1}{10},\nicefrac{1}{100})$
and measurement error $\sim N(0,\nicefrac{1}{10})$.\label{fig:lme}}
\par\end{centering}
\noindent \centering{}\subfloat[Generated trajectories.]{\noindent \centering{}\includegraphics{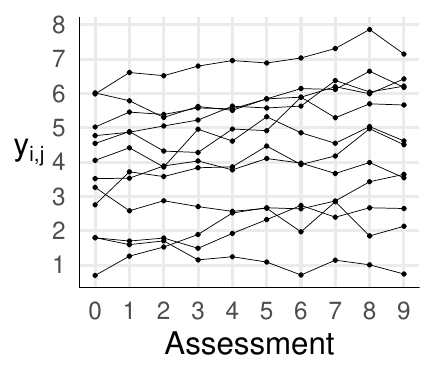}}\subfloat[Estimated multilevel model.]{\noindent \centering{}\includegraphics{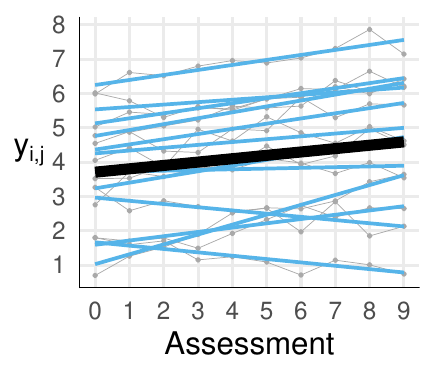}}
\end{figure}

A possible way of modeling longitudinal change is to incorporate time
as a covariate into the model. First- or second-order polynomials
are commonly used to describe change as a function of time. If more
flexible curves are required, cubic splines or fractional polynomials
can be used. Figure \ref{fig:lme} illustrates a first-order linear
mixed model describing the outcome of each individual over time by
an intercept and slope, where the assessment represents time.

In linear mixed-effects modeling, the response is assumed to be normally
distributed (although extensions exist \citep{fitzmaurice2011longitudinal}),
and the fixed and random effects are assumed to be a linear combination
of covariates, giving
\begin{align}
y_{i,j} & =\boldsymbol{x}_{i,j}\boldsymbol{\beta}+\boldsymbol{z}_{i,j}\boldsymbol{u}_{i}+\varepsilon_{i,j}\nonumber \\
\boldsymbol{u}_{i} & \sim N(0,\Sigma)\label{eq:lmm}\\
\varepsilon_{i,j} & \sim N(0,\sigma_{\varepsilon}^{2}).\nonumber 
\end{align}
Here, $\boldsymbol{x}_{i,j}$ denotes the patient-specific covariates
at time $t_{i,j}$, and $\boldsymbol{\beta}$ are the respective coefficients.
The random effects design vector is denoted by $\boldsymbol{z}_{i,j}$,
where the random effects $\boldsymbol{u}_{i}$ are jointly normally
distributed with zero mean and variance-covariance matrix $\Sigma$,
and uncorrelated with $\varepsilon_{i,j}$. The measurement error
is denoted by $\varepsilon_{i,j}$ and is assumed to be independently
normally distributed with zero mean, a common variance $\sigma_{\varepsilon}^{2}$,
and uncorrelated. Alternatively, the residuals can be modeled to be
serially correlated (autocorrelated), but more complex correlation
structures are often not possible with the inclusion of random effects
due to identifiability problems.

Mixed modeling is advantageous over fitting individual regression
models especially for datasets with a small number of measurements
per trajectory, as the estimates of the subject-specific trajectory
coefficients are more reliable due to the partial pooling of information
across subjects. Mixed effects models can be estimated with maximum
likelihood estimation. Alternatively, a Bayesian sampling approach
can be taken. This has the advantage that researchers are able to
incorporate domain knowledge in each model parameter, improving model
estimation especially under small sample size due to the ability to
include of prior knowledge \citep{spiegelhalter1994bayesian}.

\subsection{Meaning of clusters}

A cluster analysis is generally exploratory in nature, meaning that
the definitions of the clusters, or even the number of clusters, are
unknown and need to be estimated from the data. There are two possible
objectives to clustering, which determines how the resulting clusters
are interpreted. In most cases, the motivation for clustering comes
from the knowledge or expectation of considerable heterogeneity. In
an indirect application of clustering, clustering is used as a tool
for approximating a heterogeneous population in terms of a finite
number of groups without any distributional assumption on the heterogeneity.
The identified subgroups may help in accounting for correlations between
longitudinal characteristics (e.g., the association between intercept
and change over time). This is applicable when the population heterogeneity
cannot be adequately modeled using a parametric approach such as multilevel
modeling. Even in the case where a parametric model can represent
the heterogeneity, clustering may be preferred as this representation
of the heterogeneity can be easier to interpret \citep{sterba2012factors,rights2016relationship}.

An alternative reason for clustering is to test or develop theories
on subgroups \citep{moffitt2003adolescence}, referred to as a direct
application of clustering. Here, the resulting clusters are regarded
as representing distinct population groups. Throughout the years however,
the approach has been criticized for the lack of a formal test or
validation of results \citep{bauer2003distributional,bauer2007observations}.
Overall, a direct application is only advisable under strong theoretic
assumptions or highly distinct (i.e., separated) subgroups. Ideally,
the clusters are defined from theory, where clustering is applied
as a confirmatory analysis, serving as an empirical validation \citep{sher2011alcohol}.
In all other cases, an indirect application is a more practical and
lenient interpretation, and is therefore generally preferred \citep{nagin2010groupclinical,sher2011alcohol,skardhamar2010distinguishing}.

\begin{figure}
\noindent \begin{centering}
\caption{\label{fig:mixdistr}Representation of a heterogeneous distribution
using different approaches. The vertical gray lines in (b), (c) and
(d) denote the class centers.}
\subfloat[The heterogeneous distribution.]{

\includegraphics{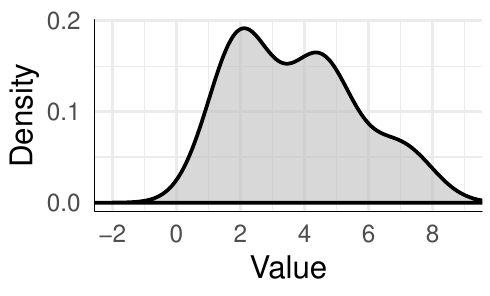}}
\par\end{centering}
\noindent \centering{}\subfloat[Representing (a) using a normal distribution.]{\includegraphics{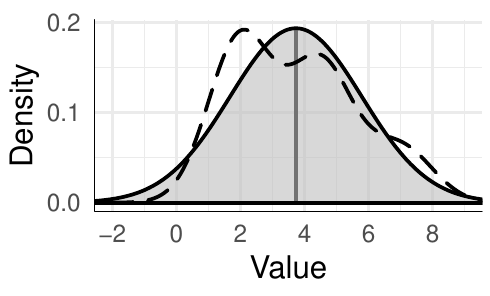}}\hfill{}\subfloat[Segmented representation of (a) using seven bins.]{\noindent \begin{centering}
\par\end{centering}
\includegraphics{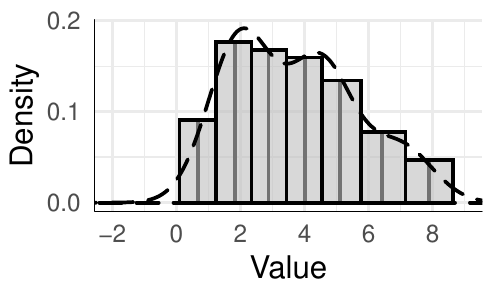}}\hfill{}\subfloat[Representing (a) using a mixture of three normals.]{

\includegraphics{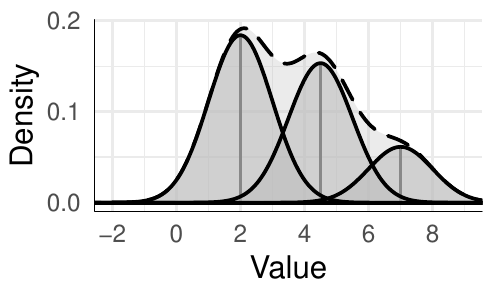}}
\end{figure}

The challenge of accounting for heterogeneity also applies to the
cluster models. An intuitive approach to clustering involves describing
the heterogeneity in terms of a number of homogeneous subgroups\@.
In contrast, modeling heterogeneity within clusters allows for a more
flexible representation of the overall heterogeneity. We illustrate
the concept visually in Figure \ref{fig:mixdistr}, depicting a heterogeneous
population in which each subject is represented by a random variable
in Figure \ref{fig:mixdistr}a. The parametric approach, assuming
a normal distribution, is shown in Figure \ref{fig:mixdistr}b.

Applying a cluster algorithm that models homogeneous clusters produces
non-overlapping bins (i.e., the clusters), represent a part of the
heterogeneity, without any assumption on the variability within the
cluster. This is illustrated in Figure \ref{fig:mixdistr}c, where
seven bins are used to represent the population density over the different
values. Due to the lack of overlap between classes, this segmentation
arbitrarily improves the approximation of the true distribution for
an increasing number of bins.

Alternatively, a parametric model can be assumed for the heterogeneity
within each cluster. Such a model represents a mixture of distributions,
referred to as a finite mixture model \citep{mclachlan2000finite}.
Figure \ref{fig:mixdistr}d shows the density of the three normal
distributions that make up the mixture model. In this example, this
was the true model from which the data was generated. This approach
has the advantage of requiring fewer classes due to the ability to
model outliers, but these models are more challenging to specify and
identify. Moreover, the overlap between classes increases as the number
of classes increases.

\section{Methods\label{sec:Methods}}

We have organized the methods for longitudinal clustering into three
approaches, with increasing model complexity. In the first approach,
a cross-sectional cluster algorithm is directly applied to the observations.
The second approach comprises feature-based estimation methods which
model the trajectories independently, and cluster the trajectories
by the model representation. The third approach involves the use of
a mixture model to perform clustering using parametric group models.
The methods are described under the assumption that the number of
clusters is part of the model specification.

\subsection{Cross-sectional clustering}

In a cross-sectional clustering approach, cluster algorithms or mixture
methods that are ordinarily applied to cross-sectional data with different
variables are applied directly to the longitudinal observations. Here,
the trajectories (or objects, in a cross-sectional context) are represented
by a sequence of observations measured at fixed times $t_{1},\ldots,t_{n}$,
with $\boldsymbol{y}_{i}=\left(y_{i,1},\ldots,y_{i,n}\right)$. Thus,
each assessment moment $t_{j}$ represents a separate (random) variable.
This representation requires individuals to be measured at (almost)
identical assessment times across subjects, although the assessment
times need not be equidistant.

Considering that cross-sectional methods do not model dependence between
parameters, applying cross-sectional methods to longitudinal data
carries the assumption that the observations are locally independent
(i.e., the temporal ordering of the observations can be ignored).
Although this assumption does not hold in a longitudinal setting,
it results in a nonparametric trajectory model that can model sudden
changes over time. The approach is therefore especially useful in
an exploratory analysis in case where the shape of the cluster trajectories
is unknown. Another reason why these methods are favorable for an
initial exploration is that they are orders of magnitude faster to
compute compared to more complex longitudinal models. The approach
is also referred to as raw data-based approach \citep{Liao2005}.
While the approach is versatile, its applicability is limited to complete
data with identical assessment times across subjects, which are challenging
requirements in case of ILD. We describe two commonly used methods
for cross-sectional clustering of longitudinal data below. The methods
are available in most statistical software packages (e.g., in \textsc{SPSS,
SAS}, \textsc{Stata}, and \textsc{R}), and have been used in practice.

\subsubsection{$k$-means clustering}

The aim of $k$-means clustering is to represent a set of objects
in terms of a predefined number of representative objects \citep{macqueen1967some}.
It is essentially a quantization method, and it is used in many fields,
including machine learning, image processing, and signal coding. In
the analysis of longitudinal data, the methodology is referred to
as $k$-means for longitudinal data (KML), or longitudinal $k$-means
analysis (LKMA). An early example of this type of analysis can be
found in the work of \citet{gude2000more}, who performed a thorough
longitudinal cluster analysis on patients with personality disorders
receiving treatment to identify groups of patients with different
symptom distress over time. The trajectories comprised three assessments
of global symptom distress. They assessed the cluster agreements between
different random starting positions, and performed post-hoc analyses
on the clusters which revealed correlations on other aspects of the
patients. Their work has been replicated recently by \citet{jensen2014heterogeneity},
with similar results. KML has been used to identify adherence patterns
in obstructive sleep apnea patients undergoing nasal CPAP therapy
\citep{wang2015pre}. Furthermore they investigated the early prediction
of the (ordered) adherence clusters using a cumulative logit model.
ANOVA F-tests were used to identify predictor variables that could
aid in predicting the adherence pattern. The KML methodology is implemented
in the R package \textit{kml}\footnote{\url{https://CRAN.R-project.org/package=kml}},
created by \citet{genolini2015kml}.

\begin{figure}
\caption{\label{fig:kmeans_example}Example of $k$-means applied to 2D data
comprised of three Gaussian clusters with means for $(y_{1},y_{2})$
of (0,0) for group A, (5,0) for group B, and (0,3) for group C, with
unit variance. The crosses denote the cluster centroids.}

\centering{}\includegraphics{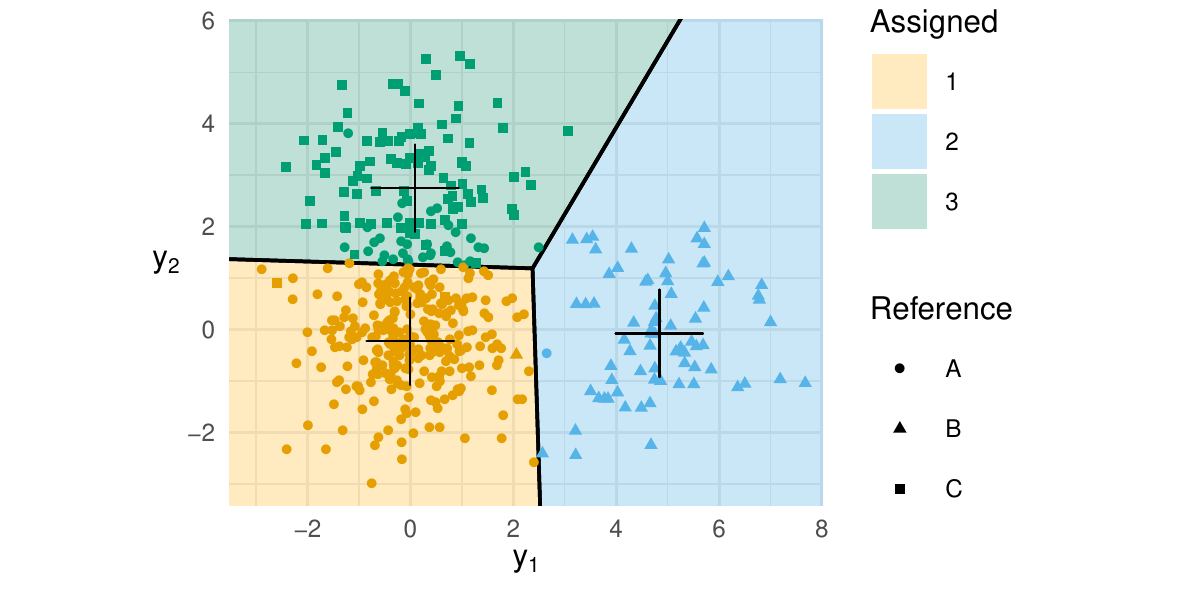}
\end{figure}

In $k$-means, clusters are assumed to be homogeneous, as each representative
object defines the center of a cluster, referred to as the centroid.
The cluster membership of objects is determined by their nearest centroid.
An example of $k$-means on synthetic 2D data is given in Figure \ref{fig:kmeans_example}.
Assuming that the total variance consists of a within-cluster and
between-cluster variance component, minimizing the within-cluster
variance ensures maximal separation of the clusters. Thus, the $k$-means
algorithm searches for the clustering $\left\{ I_{1},I_{2},...,I_{G}\right\} $
that minimizes the within-cluster sum of squares, with each cluster
$I_{g}$ having one or more objects. The objective function is described
by

\begin{equation}
\underset{I_{1},\ldots,I_{G}}{\arg\min}\sum_{g=1}^{G}\sum_{i\in I_{g}}||\boldsymbol{y}_{i}-\hat{\boldsymbol{\mu}}{}_{g}||^{2},
\end{equation}
where $\hat{\boldsymbol{\mu}}_{g}$ denotes the centroid of cluster
$g$. Finding the optimal cluster assignments is computationally infeasible
as it requires iterating over all possible assignment combinations
for all objects. Instead, the algorithm uses a heuristic iterative
approach. The solution is sensitive to the choice of the initial centroids.
The centroids can be selected, for example, by selecting $k$ objects
at random as the centroids \citep{macqueen1967some}, or using the
output from a cluster algorithm such as agglomerative hierarchical
clustering (as seen in the analysis by \citet{axen2011clustering}).
A method proposed by \citet{arthur2007k}, named $k$-means++, generally
provides better starting conditions by selecting a disperse set of
centroids at random.

The $k$-means method assumes that the within-cluster variance is
equal across clusters. When the subgroups in the data have different
variation, the estimated cluster boundaries will likely be wrong,
even when the centroids are estimated correctly. The challenge is
that cluster assignments can be problematic if many objects are relatively
distant from the respective cluster centroid (i.e. outliers), or being
close to the cluster boundary in-between clusters. An adaptation of
$k$-means, named fuzzy $c$-means, addresses this concern by using
probabilistic cluster assignment based on the distance to the centroids
\citep{dunn1973fuzzy,bezdek1981pattern}. Another challenge is the
presence of subject outliers, as these are not represented by the
cluster centroids. An example of this can be seen in Figure \ref{fig:mixdistr}c
\vpageref{fig:mixdistr}, where the tails of the distribution fall
outside any of the bins. In some cases, these outliers can affect
the resulting cluster centers. This can be prevented by excluding
these subjects from the data (referred to as trimmed $k$-means).

An advantage of $k$-means is that the algorithm scales well and converges
to a solution relatively quickly. In some studies, the trajectories
are stratified prior to clustering as a way to guide the clustering
process. An example of this approach is seen in the work of \citet{chen2007recovery},
where the authors evaluated patterns of change in self-reported back
pain over one year of time. The change in pain intensity over time,
as computed using linear regression, is used to stratify trajectories
in three strata (decreasing, increasing, and constant pain intensity),
and clustering is performed within the strata.

\paragraph*{Case study}

\begin{figure}
\caption{\label{fig:kml_case}KML case study analysis.}

\subfloat[\label{fig:kml_case_bic}BIC per solution (lower is better).]{\includegraphics{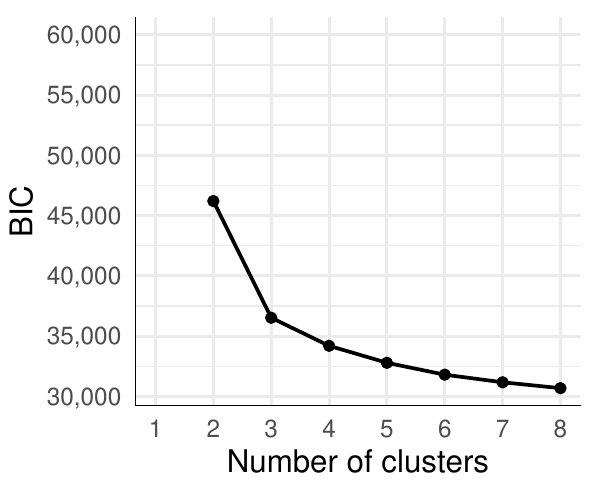}}\hfill{}\subfloat[\label{fig:kml_case_groups}The identified cluster trajectories.]{\includegraphics{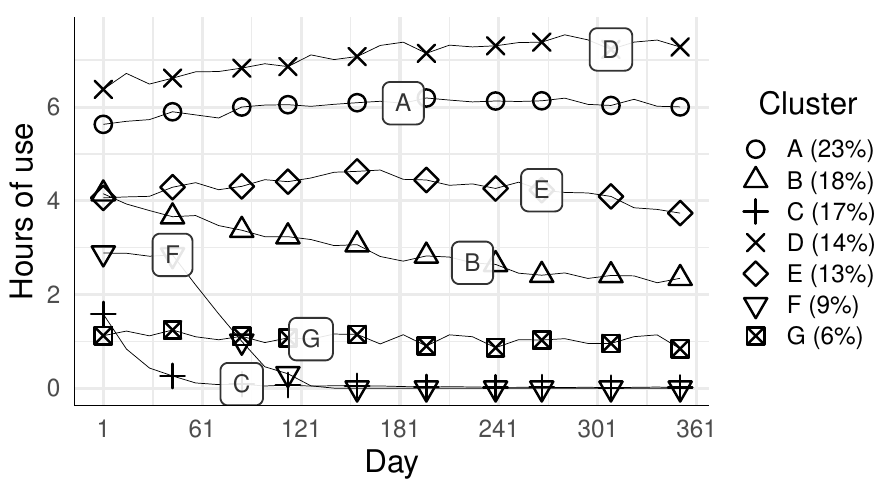}}
\end{figure}

We use the R package \textit{kml} (version 2.4.1) to cluster the
trajectories \citep{genolini2015kml}. For each number of clusters,
we run the estimation procedure 20 times, and select the best solution
from the repeated runs based on the BIC. The successive solutions
for an increasing number of clusters consistently improve the model
fit, suggesting a solution with a large number of clusters. The package
computes the BIC corresponding to the best solutions for 2 to 8 clusters,
as depicted in Figure \ref{fig:kml_case_bic}. There is a balance
to be found between the practical aspect of the number of clusters
and the improvement in model fit. Arguably, the three-cluster solution
may be preferred as the latter solutions add relatively little improvement.
However, with the objective of identifying patterns of adherence and
the improved model fit, we visually assessed the remaining solutions.

We selected the seven-cluster solution because from this solution
onward, the occasional attempters were distinguished from the non-users.
The identified cluster trajectories are shown in Figure \ref{fig:kml_case_groups}.
Although the number matches the true number of clusters, this is only
incidental, as KML failed to identify two cluster trajectory shapes
correctly, and this does not improve by introducing more clusters.
Overall, the solution recovered most of the cluster trajectories,
demonstrating the benefit of a nonparametric approach in an exploratory
setting.

\subsubsection{Latent profile analysis\label{subsec:Latent-profile-analysis}}

Latent profile analysis (LPA) is a statistical approach in which subjects
are modeled to belong to one of several unknown clusters (i.e., profiles)
\citep{lazarsfeld1968latent,vermunt2002latent}. Furthermore, the
measurement error is taken into account into the probability of belonging
to a certain class. LPA describes a mixture of profiles represented
by multivariate normal distributions, an approach also referred to
as Gaussian mixture modeling, or model-based clustering \citep{aghabozorgi2015time}.
The method is also commonly referred to as latent class analysis (LCA),
although in some fields this name specifically refers to a model involving
categorical rather than continuous observations.

Similar to KML, LPA can be applied for the identification of longitudinal
patterns without any assumption on the shape by modeling the observations
as locally independent variables at the subject level \citep{feldman2009new,twisk2012classifying}.
This type of application of LPA is sometimes specifically referred
to as a longitudinal latent-profile analysis (LLPA). The expected
value of an observation at time $t_{j}$ depends on the cluster membership.
Given the cluster membership $g$, we have

\begin{equation}
y_{i,j}=\mu_{g,j}+\varepsilon_{g,i,j},\quad i\in I_{g},\label{eq:lpa}
\end{equation}
where $\mu_{g,j}$ represents the cluster-specific mean at time $t_{j}$,
$\varepsilon_{g,i,j}\sim N(0,\sigma_{g,j})$, and $\sigma_{g,j}$
is the cluster-specific standard deviation at time $t_{j}$. The probability
density of the observations of subject $i$ is computed by marginalizing
over all $G$ clusters, giving

\begin{equation}
f\left(\boldsymbol{y}_{i}\right)=\sum_{g=1}^{G}\pi_{g}\prod_{j=1}^{n}\phi\left(y_{i,j}|\mu_{g,j},\sigma_{g,j}\right),\label{eq:lpa_dens}
\end{equation}
where $\phi(\cdot)$ denotes the probability density function of the
normal distribution, and $\pi_{g}$ denotes the cluster proportion
with $\pi_{g}>0$ and $\sum_{g=1}^{G}\pi_{g}=1$. In order to reduce
the number of parameters, the variance is commonly assumed to be equal
across measurements over time, i.e., $\sigma_{g,j}=\sigma_{g}$ \citep{peugh2013modeling}.

The model is usually estimated through maximum likelihood estimation
using the EM algorithm \citep{mclachlan2000finite}. Here, the data
is sought to be explained in terms of the unknown observation model
$\boldsymbol{\theta}=(\pi_{1},\ldots,\pi_{G-1},\boldsymbol{\mu}_{1},\ldots,\boldsymbol{\mu}_{G},\boldsymbol{\sigma}_{1},\ldots,\boldsymbol{\sigma}_{G})$,
and the unknown cluster membership matrix $z$, where $z_{i,g}$ is
the probability of patient $i$ belonging to cluster $g$ conditional
on $\boldsymbol{\theta}$. The algorithm takes an iterative approach,
involving an alternating estimation of $z$ and $\boldsymbol{\theta}$,
conditional on the other. In the E-step, the cluster assignment probabilities
are estimated from the given parameters $\boldsymbol{\theta}$ and
$\boldsymbol{y}$. In the M-step, the parameters $\boldsymbol{\theta}$
are estimated given $z$. The iterations are repeated until the improvement
in log-likelihood is sufficiently low. The estimation must be initialized
with some values for $\boldsymbol{\theta}$. Here, random values can
be used, or preferably, the output of a less complex cluster model.

While LPA is more computationally expensive than KML, it allows for
greater flexibility in fitting the data due to the ability to account
for cluster-specific variances, and even time-specific variances \citep{magidson2002latent}.
LPA is available in many software packages, including in \textsc{Mplus}\footnote{\url{https://www.statmodel.com/}}
\citep{muthen2012mplus}, \textsc{Latent GOLD}\footnote{\url{http://www.statisticalinnovations.com/latent-gold-5-1/}}
\citep{vermunt2016technical}, and the R package \textit{mclust}\footnote{\url{https://CRAN.R-project.org/package=mclust}}
\citep{scrucca2016mclust}.

\paragraph*{Case study}

\begin{figure}
\caption{\label{fig:llpa_case}LLPA case study analysis.}
\subfloat[\label{fig:llpa_case_bic}BIC per solution (lower is better).]{\includegraphics{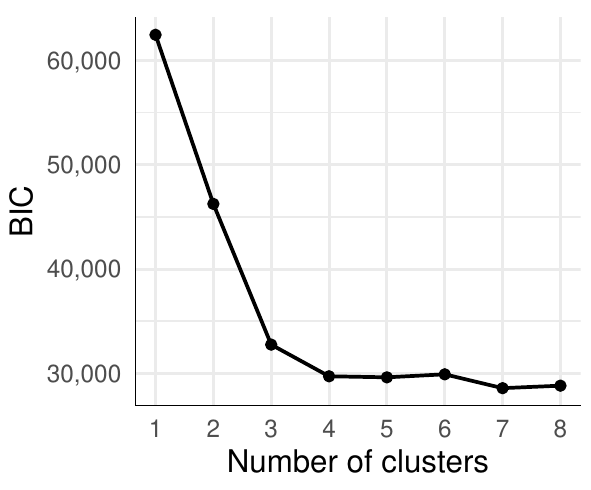}}\hfill{}\subfloat[\label{fig:llpa_case_groups}The identified cluster trajectories.]{\includegraphics{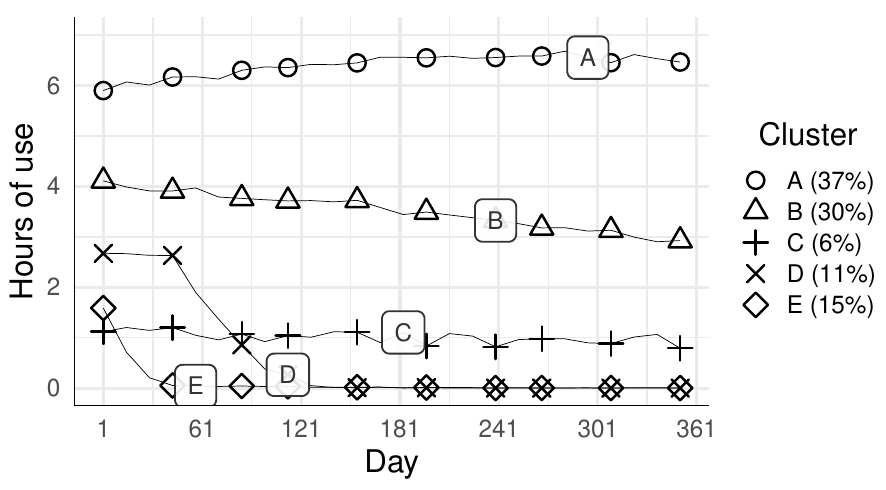}}
\end{figure}

We estimate the LPA models using the \emph{mclust} package (version
5.4.5) in R \citep{scrucca2016mclust} with cluster-specific diagonal
covariance matrices. . Ten repeated runs were found to be sufficient
in identifying the best solution per number of clusters (determined
by the BIC). The BIC per number of clusters is visualized in Figure
\ref{fig:llpa_case_bic}, showing a considerable improvement up to
four clusters. While the eight-cluster solution compares favorably,
it comprises a small clusters of only ten patients, which would limit
the power of a post-hoc analysis. Based on the BIC, one would ordinarily
select the four-cluster solution. However, upon inspection of the
successive solutions, the five-cluster solution distinguishes the
early drop-outs from the non-users, which would be of added value
in an exploratory analysis for patterns of adherence. Moreover, the
solutions involving more than five clusters comprise spurious empty
clusters, or clusters which are too small to be of practical use.

The cluster trajectories of the preferred five-cluster solution are
shown in Figure \ref{fig:llpa_case_groups}, showing an emphasis on
representing trajectories with lower usage due to the modeling of
cluster-specific and time-varying variances, because these decline
over time for the early drop-out and non-user groups.

\subsection{Distance-based clustering}

In a distance-based cluster approach, trajectories are clustered based
on their pairwise similarity, as measured by a user-specified dissimilarity
metric, i.e., distance measure. This approach comprises cluster methods
for which the user can specify an arbitrary distance metric. This
enables fast experimentation with different measures of similarity
suitable to the application at hand. The distance between two trajectories
$\boldsymbol{y}_{1}$ and $\boldsymbol{y}_{2}$ is defined by a distance
measure $d(\boldsymbol{y}_{1},\boldsymbol{y}_{2})$. A commonly used
measure is the Euclidean distance 
\begin{equation}
d(\boldsymbol{y}_{1},\boldsymbol{y}_{2})=\sqrt{\sum_{j}(y_{1,j}-y_{2,j})^{2}},
\end{equation}
which essentially yields a raw-data based approach. However, the advantage
of a distance-based approach is that domain knowledge can be taken
into account in specifying the distance measure to capture the relevant
properties of the trajectories. The Euclidean distance has been shown
to be applicable to longitudinal data \citep{genolini2010kml}, resulting
in cluster trajectories with arbitrary shapes, but conversely the
measure is sensitive to temporal offsets between subjects, and noise.
Many alternative distance measures have been suggested, including
measures that account for temporal offsets (e.g., dynamic time warping),
or reduce the complexity of the trajectory (e.g., piecewise-constant
approximation) \citep{aghabozorgi2015time,wang2013experimental}.
Another advantage is that the pairwise distances between trajectories
yields a hierarchy which provides additional information on the heterogeneity.

\subsubsection{Agglomerative hierarchical clustering}

Hierarchical clustering is a type of cluster method that identifies
a hierarchy in a set of objects based on a distance measure. The resulting
hierarchy provides an ordering of the objects based on their similarities,
which can be a useful tool in visualizing a spectrum of trajectories
with different shapes. The number of clusters can be estimated from
the distance between hierarchical clusters \citep{islam2015comparison}.

Agglomerative hierarchical clustering (AHC) uses a bottom-up approach
to identify the hierarchical structure of the objects. Each of the
objects start out as separate clusters. The AHC approach is commonly
used in combination with a post-hoc analysis to identify factors that
may differ between clusters. \citet{babbin2015identifying} investigated
the daily time on CPAP therapy of patients with obstructive sleep
apnea to identify clusters of patients with similar adherence trends.
Other examples include the investigation of \citet{hoeppner2008detecting}
of daily smoking patterns after patients went through a reduction
program, and patterns of alcohol use \citep{harrington2014typology}.

\begin{figure}
\caption{\label{fig:ahc_example}Example of a dendrogram computed from longitudinal
data comprising three groups, each having three trajectories. The
cluster trajectories are described by an intercept and slope, with
coefficients $\boldsymbol{\beta}_{A}=(3,-0.3)$, $\boldsymbol{\beta}_{B}=(2,0)$,
and $\boldsymbol{\beta}_{C}=(0,0.2)$, respectively.}

\centering{}\includegraphics{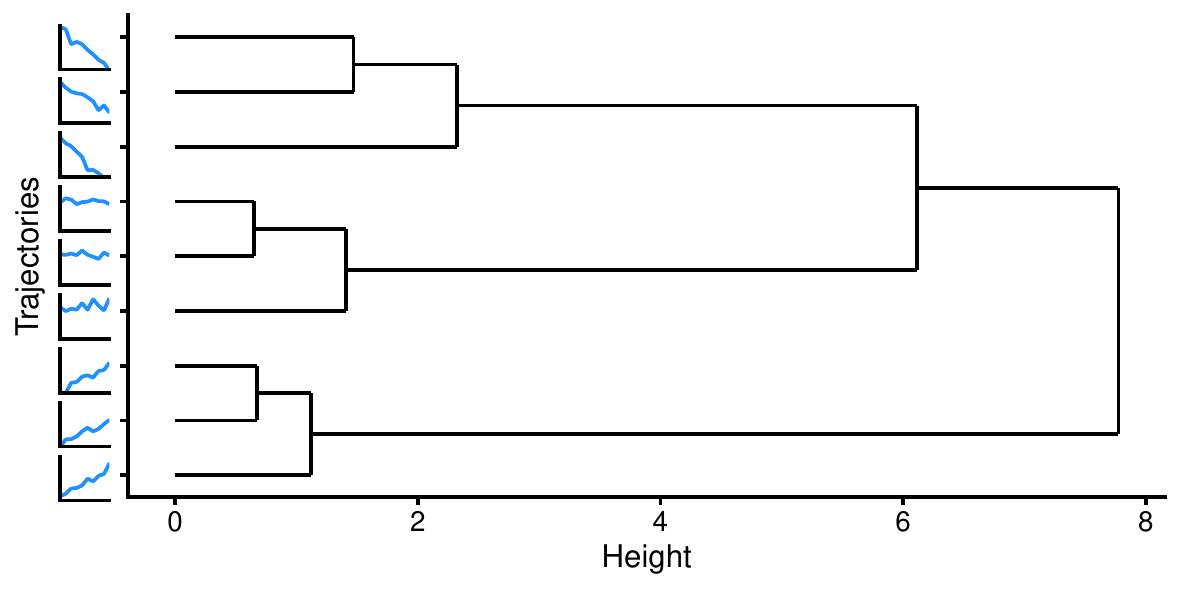}
\end{figure}

In AHC, the hierarchy is identified using a greedy approach, where
at each step the two most similar clusters are combined into a new
cluster. This is repeated until a single cluster remains containing
all objects. The resulting hierarchy can be visualized in a dendrogram.
To illustrate, Figure \ref{fig:ahc_example} depicts the hierarchy
of nine trajectories generated from three different linear models.
Combining objects and clusters involves two distance measures. Firstly,
a distance measure $d\left(\boldsymbol{y}_{i},\boldsymbol{y}_{j}\right)$
is needed for determining the similarity of the trajectory of individual
$i$ and individual $j$, with $i\neq j$. Secondly, a distance measure
between clusters $I_{r}$ and $I_{s}$ is needed, where a cluster
$I_{g}$ is the subset of individuals (out of all individuals $I$)
that belong to cluster $g$. The distance $d\left(I_{r},I_{s}\right)$
is referred to as the linkage criterion, with $r\neq s$.

An intuitive approach to measuring the distance between clusters is
to measure the average pairwise distances between the clusters. This
is referred to as the unweighted average linkage (UPGMA), and is computed
by
\[
d(I_{r},I_{s})=\frac{1}{|I_{r}|\cdot|I_{s}|}\sum_{i\in I_{r}}\sum_{j\in I_{s}}d(\boldsymbol{y}_{i},\boldsymbol{y}_{j}).
\]
Alternative linkage criteria which are commonly used include the minimum
linkage $\min\{d(\boldsymbol{y}_{i},\boldsymbol{y}_{j}):i\in I_{r},j\in I_{s}\}$,
centroid linkage, and Ward's minimum variance method.

AHC provides a good trade-off between finding a reasonable hierarchy
quickly, and identifying the optimal hierarchy (i.e., the hierarchy
that minimizes the overall distance). However, the computation time
quickly grows with the number of trajectories, due to the pairwise
distances that must be computed between all subjects.

\paragraph*{Case study}

\begin{figure}
\noindent \begin{centering}
\caption{\label{fig:ahc_case}Case study analysis using AHC.}
\par\end{centering}
\noindent \begin{centering}
\subfloat[\label{fig:ahc_case_tree}The identified trajectory hierarchy.]{\noindent \centering{}\includegraphics{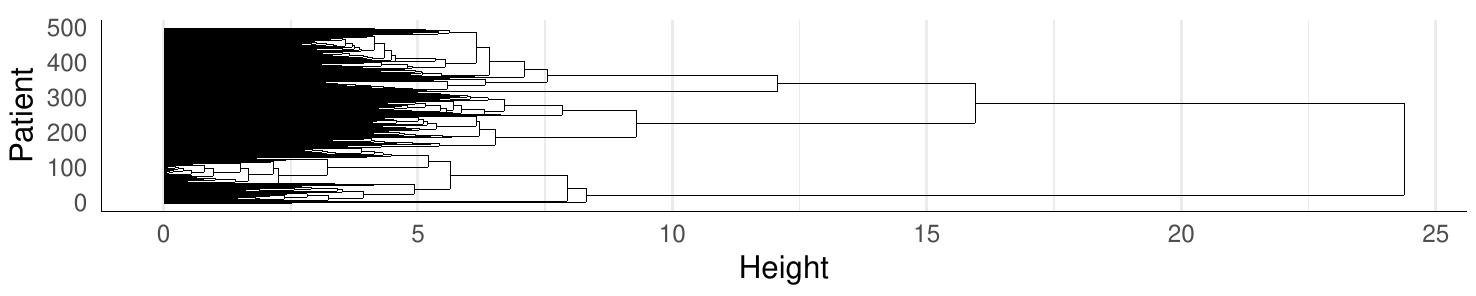}}
\par\end{centering}
\noindent \centering{}\subfloat[\label{fig:ahc_case_sil}Average silhouette width per solution (higher
is better).]{\noindent \includegraphics{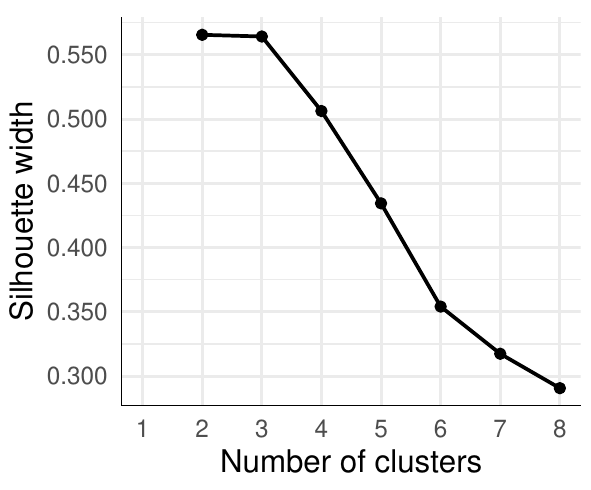}}\hfill{}\subfloat[\label{fig:ahc_case_groups}The identified cluster trajectories.]{\noindent \includegraphics{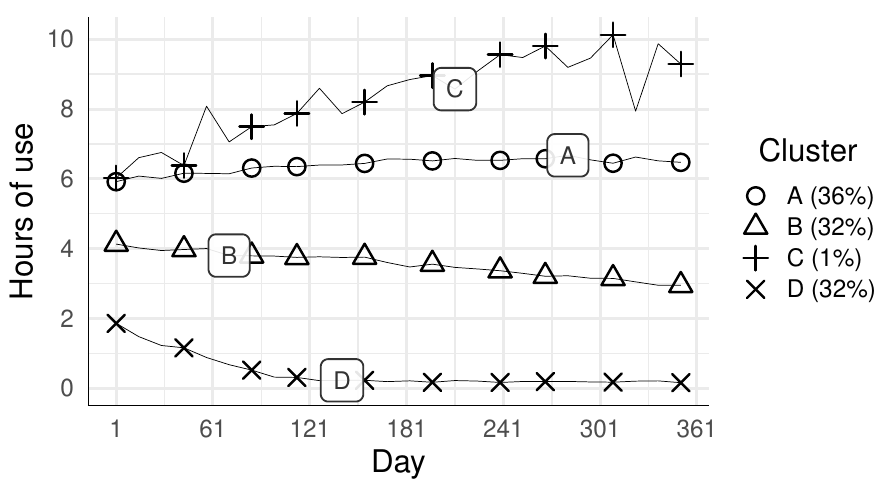}}
\end{figure}

As all measurements across patients are aligned in the case study,
we can apply the Euclidean distance to compute the pairwise similarity
between patient trajectories. We then apply the agglomerative hierarchical
cluster algorithm that is available in R using the average linkage.
For each number of clusters, the trajectory assignments are obtained
based on the identified hierarchy visualized in Figure \ref{fig:ahc_case_tree}.
The solutions are compared using the ASW. As can be seen in Figure
\ref{fig:ahc_case_sil}, the ASW is considerably lower for solutions
with more than three clusters. A cluster solution with an ASW above
0.5 is generally considered to have some consistent structure. In
an exploratory setting it may be worthwhile to forfeit this rule of
thumb in favor of identifying additional meaningful temporal patterns,
given that the clusters are of sufficient size. In this case however,
the solutions with a larger number of clusters include clusters of
outliers comprising only a single trajectory (as can be seen from
the hierarchy), so the three-cluster solution is preferred.

The solution comprising three clusters is shown in Figure \ref{fig:ahc_case_groups}.
Here, the cluster trajectories are computed by averaging across all
trajectories that are assigned to it. The solution provides a balanced
representation of the seven groups, combined based on the respective
mean level.

\subsection{Feature-based clustering}

In a feature-based clustering approach, individual trajectories are
described in terms of a parametric model that captures the relevant
characteristics. Here, each trajectory $\boldsymbol{y}_{i}$ is reduced
to a set of model parameters $\boldsymbol{b}_{i}=\left(b_{i,1,}b_{i,2,},...,b_{i,p}\right)$
which can be regarded as the $p$ features of the trajectory. Clusters
of trajectories with similar characteristics can then be identified
by applying a cross-sectional cluster algorithm to the model parameters.
The appeal of a feature-based clustering approach is that researchers
can incorporate domain knowledge in defining the similarity between
trajectories by using an appropriate model, or by computing several
independent characteristics. The characteristics may better capture
the differences between trajectories than a cross-sectional approach
based on the shape alone, resulting in more well-separated clusters
\citep{wang2006characteristic}. The approach is also commonly referred
to as a feature-based or model-based approach\footnote{The term ``model-based clustering'' appears to be used for both
feature-based clustering of model parameters and mixture modeling.} \citep{aghabozorgi2015time}.

The second step of clustering is easy to implement and available in
most statistical software packages through the widespread availability
of clustering algorithms such as $k$-means. There are several strengths
to the approach; especially within the context of ILD. Firstly, the
parametric representation of trajectories is naturally more robust
to missing observations, as the computed characteristics tend to be
based on multiple observations. Secondly, the approach can handle
trajectories of varying lengths between individuals or measured at
different intervals \citep{wang2006characteristic}. Lastly, the method
scales well with the amount of data, as the representation is of constant
size independent of the number of observations. Moreover, the trajectory
representations only need to be computed once, after which a cluster
algorithm can be fitted to the feature data for varying settings as
part of the model selection.

\subsubsection{Individual time series representations\label{subsec:Time-series-models}}

Trajectories can be represented in many ways. An intuitive approach
to describing each trajectory is in terms of a linear model dependent
on time (e.g., an intercept and slope), as seen in multilevel modeling,
where the individual trajectory representations can be obtained from
the estimated random effects, and are then clustered using a cluster
algorithm (e.g., $k$-means \citep{twisk2012classifying}). While
this is a useful approach when there are relatively few observations
per trajectory, independently estimating the representation of each
trajectory frees researchers of any assumptions on the population
heterogeneity, providing a better fit \citep{liu2017person}. This
approach is referred to as an individual time series (ITS) analysis
\citep{bushway2009measuring,greenberg2016criminal,liu2017person}.

An example of the ITS approach to clustering is seen in a method named
anchored $k$-medoids, created by \citet{adepeju2019akmedoids}, where
the trajectories are represented by time-dependent linear regression
models. The trajectories are then clustered based on the coefficients
using $k$-medoids. The $k$-medoids cluster algorithm is similar
to $k$-means, but uses one of the observations (i.e., objects) as
the cluster center instead of an average across observations. This
is especially useful for ITS representations because the algorithm
can handle arbitrary distance metrics, and does not require the computation
of an average cluster representation, which may not be sensible for
some model coefficients or distance metrics. 

There are two advantages to the ITS approach. Firstly, the trajectory
models can be estimated independently, allowing for a trivial parallelization
of the estimation process. Secondly, the independent models do not
need to account for any variability between trajectories, and are
therefore easier to estimate than a multilevel approach. A disadvantage
of modeling each trajectory independently is that there may be trajectories
for which the model fit is poor, resulting in clusters primarily containing
poorly fitted models of which the original trajectories may not be
similar. A poor fit can be the result of a trajectory not meeting
the model assumptions, or the number of data points being insufficient
for the model. The possibility to combine multiple representations
into a single model vector provides additional challenges similar
to those seen in cross-sectional clustering involving high-dimensional
data: The coefficients may need to be normalized to ensure that the
distance function is not disproportionately affected by coefficients
of a higher magnitude. On the other hand, a subset of coefficients
may be deemed more important \citep{fulcher2014highly}.

In its simplest form, trajectories can be represented by summary statistics
such as the mean, standard deviation, skewness, range, degree of stationarity,
periodicity, autocorrelation, or entropy \citep{fulcher2013highly,aghabozorgi2015time}.
Another practical approach is to categorize the response variable
into a finite number of values (i.e., states). \citet{kiwuwa2011clustering}
modeled the adherence behavior of patients undergoing medical treatment
for HIV infection using a first-order Markov chain, modeling the transitional
probabilities of the non-adherent and adherence states. They used
AHC using Ward's minimum variance method to cluster the transitional
probability vectors. A limitation of many of the statistical measures
proposed is that they are under the assumption that the statistical
properties do not change over time. This can be resolved by correcting
for longitudinal changes, by estimating the properties over multiple
segments of the trajectory, or by fitting a linear model that represents
the change over time.

In other cases, an abrupt change in the observations may be expected
from domain knowledge. Change points are for example modeled in the
work of \citet{axen2011clustering}, who investigated patients diagnosed
with non-specific low back pain. In this work, the trajectories were
modeled using two linear models which describe the early and late
trajectory, respectively, fitted using spline regression. The linear
model coefficients, as well as the estimated intersection point of
the two lines, were used as inputs for the second step clustering.

\citet{wang2006characteristic} propose a set of nine statistical
features for describing a trajectory: Firstly, a trajectory can be
decomposed into several components \citep{kendall1983advanced}; a
trend $T_{t}$ (the long-term average level), a seasonal effect $S_{t}$,
and a cyclic effect $C_{t}$ (also referred to as periodicity or frequency).
Assuming that the components are not proportional, a time series can
be described using an additive model
\begin{equation}
y_{t}=T_{t}+C_{t}+S_{t}+\varepsilon_{t},
\end{equation}
where $\varepsilon_{t}$ denotes the irregular component (i.e., the
residual). Secondly, Wang et al. suggest to describe aspects of the
measurement distribution in terms of the skewness and kurtosis. Thirdly,
the temporal structure of the data is expressed in terms of the autocorrelation
and a test for non-linearity, e.g., through a nonparametric kernel
test or neural network test. Lastly, the trajectory complexity is
assessed using self-similarity (a measure of long-term dependence)
and other methods commonly used in describing chaotic systems (e.g.,
the Lyapunov exponent, which is a measure of divergence in response
to small perturbations). Especially, the latter metrics require a
sizable number of observations per trajectory to be estimated reliably,
so these are only suitable for ILD.

The irregular component $\varepsilon_{t}$ describes the local changes
of a trajectory. A straightforward way to describe the component is
through a white noise process of zero mean and variance $\sigma^{2}$,
assuming independent and identically distributed observations. When
the local changes are assumed to correlate with past values, an autoregressive
(AR) model is typically used. This model regresses past values using
a $p$\textsuperscript{th}-order polynomial. Alternatively, the model
error may depend on previous errors, which can be described using
a moving average (MA) model of the past $q$ error terms. Combining
these two models, we obtain an ARMA$(p,q)$ model describing a stochastic
process
\begin{align}
y_{t} & =c+\sum_{i=1}^{p}\phi_{i}y_{t-i}+\sum_{j=1}^{q}\theta_{j}\epsilon_{t-j}+\epsilon_{t},\label{eq:arma}
\end{align}
with $c$ describing the model intercept, and $\phi_{1,...,p}$ and
$\theta_{1,...,q}$ describing the parameters of the AR and MA models,
respectively. The model residuals are denoted by $\epsilon_{t}$,
and are assumed independent and follow a normal distribution with
zero mean and variance $\sigma_{\epsilon}^{2}$. ARMA can be applied
to non-stationary data by first applying one or more differencing
steps $y_{t}^{\prime}=y_{t}-y_{t-1}$ to the data, in which case the
approach is referred to as ARIMA (where the I stands for integrated).
ARIMA is mostly used in the economical and financial domain for predicting
future values, but it is also of use for process modeling (e.g., adaptive
control), and descriptive modeling \citep{jebb2015time,aloia2008time}.
\citet{kalpakis2001distance} have proposed a distance measure for
comparing ARIMA models, which are then clustered using $k$-medoids.
This approach could be regarded as a hybrid of the feature-based and
distance-based approaches to longitudinal clustering. Other useful
methods for describing stochastic processes are autoregressive conditional
heteroskedasticity (ARCH), Gaussian processes, and state space models
\citep{fulcher2013highly}.

\paragraph*{Case study}

We model each patient independently on several aspects. Each trajectory
is represented in terms of an intercept $\beta_{i,0}$, an orthogonal
polynomial of degree two with coefficients $\left(\beta_{i,1},\beta_{i,2}\right)$,
a residual error $\sigma_{\varepsilon,i}$, and the log-number of
attempted days $\log N_{i}$. This yields the patient representation
$\boldsymbol{b}_{i}=($$b_{i,1}=\beta_{i,0}$, $b_{i,2}=\beta_{i,1}$,
$b_{i,3}=\beta_{i,2}$, $b_{i,3}=\sigma_{\varepsilon,i}$, $b_{i,4}=\log N_{i})$.
The patient representation vectors $\boldsymbol{b}_{i}$ are scaled
to ensure zero mean and unit variance across the features. We compute
a distance matrix using the Euclidean distance, and then apply $k$-medoids
using the \textit{cluster} package\footnote{\url{https://CRAN.R-project.org/package=flexmix}}
(version 2.1.0) \citep{maechler2019cluster} in order to obtain clusters
which are represented by one of the computed representation vectors.
Similar to the AHC analysis, we evaluate cluster solutions using the
ASW.

The inclusion of irrelevant features can negatively affect the cluster
separation. It is therefore important to select the relevant aspects.
Moreover, the approach is sensitive to spurious estimates of the patient-specific
coefficients, as only a limited number of observations are available.
These aspects reduce the separation between clusters, resulting in
a lower ASW. We investigated different subsets of the patient representation
vector by assessing the highest observed ASW. This revealed that the
inclusion of $b_{i,3}=\sigma_{\varepsilon,i}$ negatively affected
cluster separation, and should be excluded. This is despite the fact
that the data was generated with some degree of group-specific variance.
It was found that the residual error is often underestimated, likely
resulting from overfitting of the polynomial trajectory of some of
the patients.

\begin{figure}
\caption{\label{fig:twostep_case}Case study analysis using feature-based clustering.}
\subfloat[\label{fig:twostep_case_sil}Average silhouette width per solution
(higher is better).]{\includegraphics{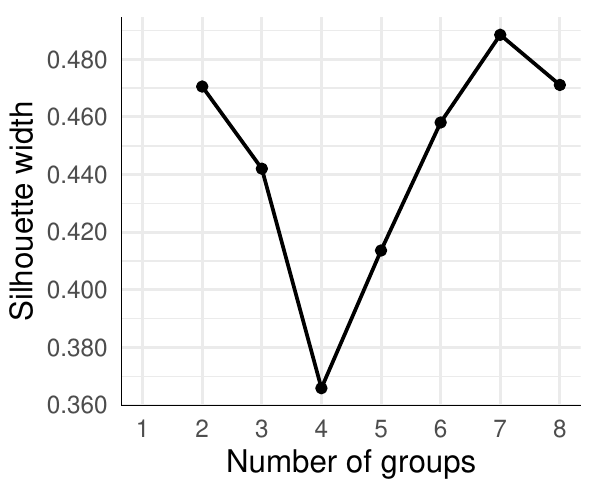}}\hfill{}\subfloat[\label{fig:twostep_case_groups}The identified cluster trajectories.]{\includegraphics{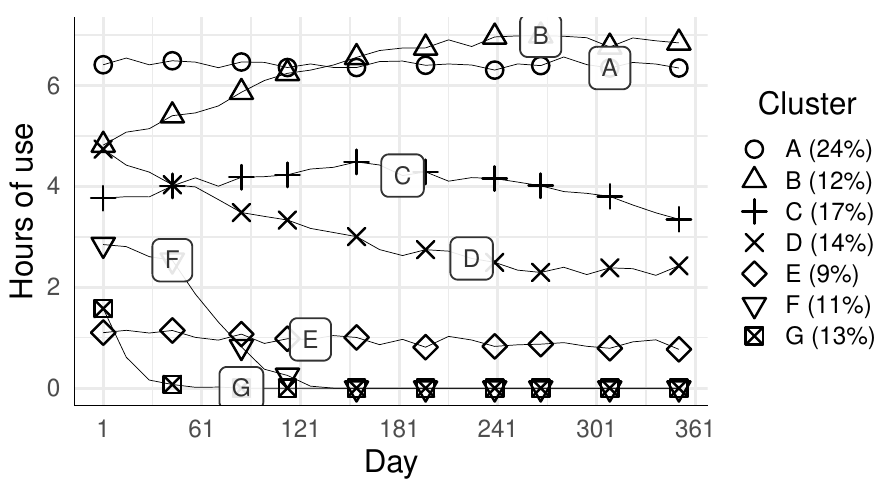}}
\end{figure}

The ASW per number of clusters for the final model $\boldsymbol{b}_{i}=\left(b_{i,1}=\beta_{i,0},b_{i,2}=\beta_{i,1},b_{i,3}=\beta_{i,2},b_{i,3}=\log N_{i}\right)$
is displayed in Figure \ref{fig:twostep_case_sil}. The highest ASW
of 0.49 is obtained for the seven-cluster solution. The cluster trajectories
visualized in Figure \ref{fig:twostep_case_groups} were obtained
by averaging across the respective trajectories. The solution matches
the ground truth, demonstrating the ability to recover the underlying
clusters when the relevant longitudinal aspects are used.

\subsection{Mixture modeling}

Mixture models describe a distribution in terms of a set of underlying
distributions, under the assumption that the distribution comprises
different data-generating processes or random variables. Usually,
the submodels assume the same parametric distribution, but with different
coefficients. An example of a mixture distribution comprising normals
was shown in Figure \ref{fig:mixdistr}d \vpageref{fig:mixdistr}.
In a statistical analysis setting, mixture models comprise a set of
regression models. Here too, the submodels tend to have an identical
specification.

The basic idea behind a longitudinal mixture model is to fit a mixture
distribution to the longitudinal observations $\boldsymbol{y}_{i}$.
Thus the mixture model density $f\left(\boldsymbol{y}_{i}|\boldsymbol{\theta}\right)$
with model parameters $\boldsymbol{\theta}=\left(\boldsymbol{\pi},\boldsymbol{\theta}_{1},\ldots,\boldsymbol{\theta}_{G}\right)$
is defined by
\begin{equation}
f\left(\boldsymbol{y}_{i}|\boldsymbol{\theta}\right)=\sum_{g=1}^{G}\pi_{g}f\left(\boldsymbol{y}_{i}|\boldsymbol{\theta}_{g}\right).
\end{equation}
Here, $f\left(\boldsymbol{y}_{i}|\boldsymbol{\theta}_{g}\right)$
denotes the conditional density of $\boldsymbol{y}_{i}$ given that
$i$ belongs to cluster $g$. The cluster membership of individual
trajectories is unknown and therefore treated as being probabilistic.
The probability of a random subject belonging to cluster $g$ is denoted
by $\pi_{g}$, where $0\leq\pi_{g}\leq1$ and $\sum_{g}\pi_{g}=1$.
This can also be interpreted as the cluster proportion. The posterior
probability of a subject belonging to a given cluster is computed
by normalizing the respective cluster density over all clusters by
\begin{equation}
\Pr(\boldsymbol{y}_{i}|i\in I_{g},\boldsymbol{\theta})=\frac{\pi_{g}f\left(\boldsymbol{y}_{i}|\boldsymbol{\theta}_{g}\right)}{\sum_{g'=1}^{G}\pi_{g'}f\left(\boldsymbol{y}_{i}|\boldsymbol{\theta}_{g'}\right)}.\label{eq:postprob}
\end{equation}
While the cluster assignments are probabilistic, meaning that a subject
can belong to any cluster with a certain probability, the subject
is usually assumed to belong to the cluster with the highest posterior
probability.

The effect of baseline covariates on the cluster membership can be
explored by including a multinomial logistic regression component
for $\pi_{g}$. For a vector of covariates $\boldsymbol{x}_{i}$ of
subject $i$, the cluster membership probability is computed by

\begin{equation}
\pi_{g}(\boldsymbol{x}_{i})=\frac{\exp(\boldsymbol{\eta}_{g}\boldsymbol{x}_{i})}{\sum_{g'=1}^{G}\exp(\boldsymbol{\eta}_{g'}\boldsymbol{x}_{i})},\label{eq:postprob_cov}
\end{equation}
where $\boldsymbol{\eta}_{g}$ denotes the multinomial regression
coefficients for cluster $g$. For the purpose of model identifiability,
the last cluster is used as the reference, with $\boldsymbol{\eta}_{G}=0$.
In the remainder of the overview, we assume a model without cluster
membership covariates for brevity.

The estimation of these mixture models follow the same approach as
the EM algorithm described in subsection \ref{subsec:Latent-profile-analysis}
on LPA, as this is a type of mixture model as well. The important
distinction is that the mixture models presented in this section allow
for an arbitrary number of measurements per trajectory, and at arbitrary
moments in time.

\subsubsection{Group-based trajectory modeling}

Similar to the concept of methods such as $k$-means or LPA, group-based
trajectory modeling (GBTM\footnote{Abbreviated as GTM in some articles.})
describes the population heterogeneity via a set of homogeneous clusters,
where subjects are only represented by their respective cluster trajectory
\citep{nagin2010groupclinical,nagin2005developmental}. In contrast,
GBTM represents the trajectories using a parametric model. It can
be regarded as a multilevel model with nonparametric random effects
(i.e., a finite number of random effect values, representing the clusters),
which is especially useful when random effects are non-normal or correlated
\citep{rights2016relationship}. The model is easy to interpret due
to its distinct cluster trajectories. The method is also referred
to as latent-class growth analysis or modeling (LCGA, LCGM), semi-parametric
group-based modeling (SGBM), TRAJ\footnote{Named after the macro in SAS, \textsc{PROC TRAJ}.},
and sometimes as nonparametric multilevel mixture modeling (NPMM).

The method originates from the field of criminology. Two decades ago,
\citet{nagin1993age} suggested a model for describing developmental
trajectories in individuals for whom the number of yearly crimes was
measured in relation to age. They proposed a longitudinal Poisson
mixture model for separating the trajectories, comprising count data,
into distinct clusters. In a later paper, \citet{nagin1999analyzing}
presented GBTM as a flexible method for identifying distinct trajectories
in a set of longitudinal measurements. Furthermore, models were proposed
that assume (censored) normal data or binary data for the observations.
Its applications extend further than the domain it was originally
created for. GBTM has been applied in the field of psychology, medicine
\citep{nagin2010groupclinical,franklin2013group}, and ecology \citep{matthews2015group},
among others.

A GBTM describes the trajectories using a linear model. The design
vector at time $t_{i,j}$ is denoted by $\boldsymbol{x}_{i,j}$. The
cluster trajectories are often modeled using polynomials. As an example,
$\boldsymbol{x}_{i,j}=\left(1,t_{i,j},t_{i,j}^{2}\right)$ describes
a second-order polynomial time trajectory. The trajectories as modeled
by a GBTM, given membership to a specific cluster $g$, are described
by
\begin{equation}
y_{i,j}=\boldsymbol{x}_{i,j}\boldsymbol{\beta}_{g}+\varepsilon_{g,i,j},\quad i\in I_{g},
\end{equation}
where $\boldsymbol{\beta}_{g}$ denotes the cluster-specific regression
coefficients, and the residual error $\varepsilon_{g,i,j}$ is assumed
to be independently normally distributed with zero mean and variance
$\sigma_{g}^{2}$. The marginal mean is computed by

\begin{equation}
\mathbb{E}(y_{i,j})=\sum_{g=1}^{G}\pi_{g}\boldsymbol{x}_{i,j}\boldsymbol{\beta}_{g}.
\end{equation}

The GBTM parameters and clusters are estimated by maximizing the likelihood
of the model for a given number of clusters $G$ using the EM algorithm.
Missing observations tend to be assumed missing at random and are
therefore ignored. The model can be adapted to fit a wide variety
of response distributions. It has also been used to model data under
a censored normal, zero-inflated Poisson, logistic, or beta distribution
\citep{jones2007advances,elmer2018using}.

\citet{jones2007advances} proposed the estimation of confidence intervals
on cluster membership probabilities and trajectories using Taylor-series
expansion. \citet{nielsen2014group} proposed an alternative to model
estimation and selection using a cross-validation error methodology.
\citet{nagin2018group} extended the GBTM to account for multiple
outcome trajectories, in which the outcomes are assumed to be conditionally
independent. This is found to be favorable to the alternative of clustering
each outcome separately and then combining the results.

There are a couple of disadvantages to modeling trajectories through
polynomials. Firstly, the possible shapes a polynomial may represent
is limited, so the model may not be able to fit the longitudinal shape.
Secondly, higher-order polynomials tend to overfit the data or produce
spurious shapes. Researchers should therefore be careful in interpreting
the shapes in detail. As a more reliable alternative, \citet{francis2016smoothing}
proposed smoothing the cluster trajectories using a cubic B-spline.
They observed an improved model fit while allowing for more flexible
cluster trajectories.

Overall, GBTM is applicable to ILD in many aspects. The model can
handle missing data, observations measured at different times, and
estimation is relatively fast due to the low number of parameters
involved. In addition, the probabilistic nature of the model has been
shown to make it suitable for real-time prediction, where cluster
membership and the expected trajectory can be computed as new observations
become available \citep{elmer2019novel}.

Implementations of GBTM are available in \textsc{SAS}\footnote{\label{fn:proctraj}The plugin is available at \url{http://www.andrew.cmu.edu/user/bjones}}
\citep{jones2001sas}, \textsc{Stata}\textsuperscript{\ref{fn:proctraj}}\citep{jones2013note},
\textsc{Mplus}\footnote{\url{https://www.statmodel.com}} \citep{muthen2012mplus}
and \textsc{OpenMx}\footnote{\url{http://openmx.psyc.virginia.edu}}
\citep{boker2011openmx}, and in R via the \textit{lcmm}\footnote{\url{https://CRAN.R-project.org/package=lcmm}}
\citep{proust2017estimation}, \textit{crimCV}\footnote{\url{https://CRAN.R-project.org/package=crimCV}},
\textit{flexmix}\footnote{\url{https://CRAN.R-project.org/package=flexmix}}
\citep{grun2008flexmix}, or \textit{mixtools}\footnote{\url{https://CRAN.R-project.org/package=mixtools}}
\citep{benaglia2009mixtools} package, among others.

\paragraph*{Case study}

Prior to the GBTM analysis, we investigate the appropriate trajectory
representation by evaluating mixed models with different polynomial
degrees in the fixed and random effects. We normalize the 26 measurement
times by scaling the range from $[1,351]$ to $[0,1]$ for numeric
stability. The mixed models and GBTMs are estimated with the R package
\textit{lcmm} (version 1.7.8), developed by \citet{proust2017estimation}.
The model fit and variance components are reported in Table \ref{tab:lmm_case}.
The residual standard error and BIC indicate an improved fit with
a higher order polynomial. While the model with polynomial representation
of degree 3 achieves the best fit, the improvement over the second
degree model is relatively small. In consideration of the linear increase
in the number of model parameters with an increasing number of clusters,
we settle for a quadratic representation.

\begin{figure}
\caption{\label{fig:gbtm_case}GBTM case study analysis.}
\subfloat[\label{fig:gbtm_case_bic}BIC per solution (lower is better).]{\includegraphics{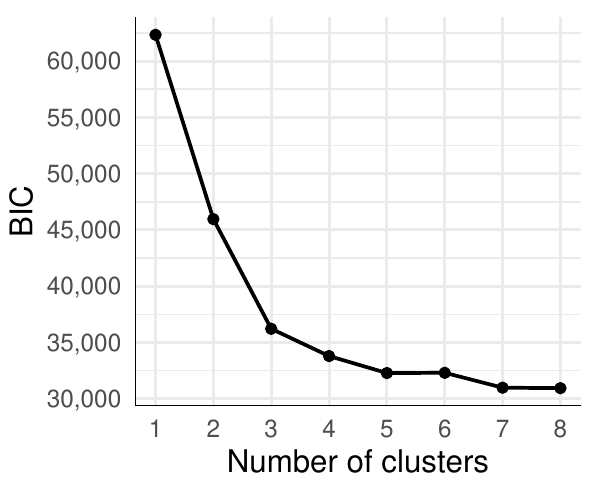}}\hfill{}\subfloat[\label{fig:gbtm_case_groups}The identified cluster trajectories.]{\includegraphics{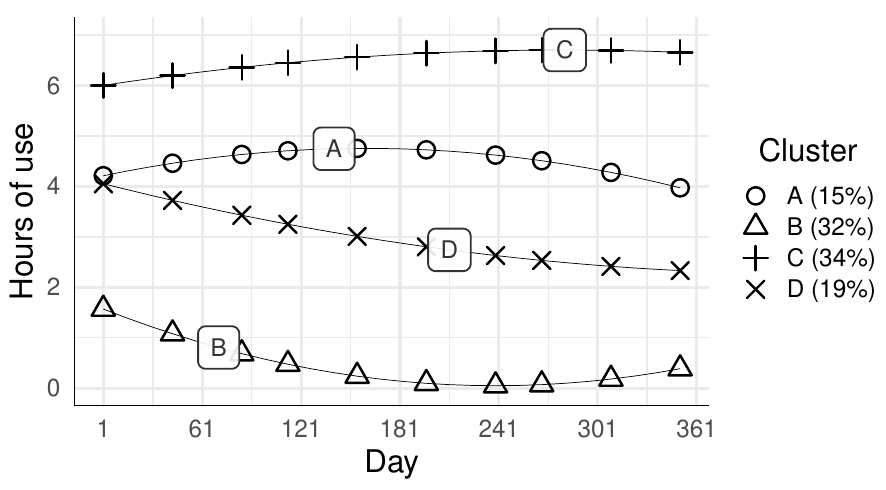}}
\end{figure}

We estimate the GBTM solutions using a grid search with 20 random
starts in order to identify a good starting position during model
optimization. As depicted in Figure \ref{fig:gbtm_case_bic}, the
model fit improves with an increasing number of clusters. Judging
from the change in improvement from one solution to the next, a three-
or four-cluster solution is preferred. Upon visual inspection of both
solutions, we opt for the four cluster solution due to the addition
of the cluster trajectory similar to the Variable users group in the
ground truth.

The four cluster trajectories are shown in Figure \ref{fig:gbtm_case_groups}.
Overall this solution adequately captures the heterogeneity of the
data. Cluster A (32\%) represents the non-users, early drop-outs,
and occasional attempters. Cluster B (35\%) comprises the slow improvers
and good users. The remaining clusters match the ground truth.
\begin{table}
\noindent \begin{centering}
\caption{\label{tab:lmm_case}Single-cluster analysis using mixed modeling
with a normalized time covariate. Here, $\sigma_{0},\ldots,\sigma_{3}$
represent the square root of the diagonal of the covariance matrix
$\Sigma$.}
\par\end{centering}
\noindent \centering{}%
\begin{tabular}{ccccccc}
Model degree & $\sigma_{0}$ & $\sigma_{1}$ & $\sigma_{2}$ & $\sigma_{3}$ & $\sigma_{\varepsilon}$ & BIC\tabularnewline
\midrule
\midrule 
0 & 2.5 & - & - & - & 0.79 & 33,433\tabularnewline
1 & 2.2 & 1.4 & - & - & 0.63 & 29,205\tabularnewline
2 & 2.1 & 4.4 & 3.2 & - & 0.58 & 27,375\tabularnewline
3 & 2.0 & 8.7 & 6.3 & 4.1 & 0.57 & 27,146\tabularnewline
\bottomrule
\end{tabular}
\end{table}

\subsubsection{Growth mixture modeling}

Growth mixture modeling (GMM) extends GBTM with the inclusion of
parametric random effects, enabling a better fit to the data under
the assumption of within-cluster variability \citep{verbeke1996linear,muthen1999finite,muthen2002general,muthen2004latent}.
The method is also described as a longitudinal latent-class mixed
model, a multilevel mixture model, or a finite mixture of mixed models.
GMM has been applied across many domains in the past decade.

Although GMM is commonly applied and described in a SEM framework,
we present it in a mixed-modeling approach in order to be consistent
with the notation of the previous sections. The longitudinal observations,
conditional on belonging to cluster $g$, can be described by the
linear mixed model specified in Equation \vref{eq:lmm} with

\begin{align}
y_{i,j} & =\boldsymbol{x}_{i,j}\boldsymbol{\beta}_{g}+\boldsymbol{\boldsymbol{z}}_{i,j}\boldsymbol{u}_{g,i}+\varepsilon_{g,i,j},\quad i\in I_{g}\nonumber \\
\boldsymbol{u}_{g,i} & \sim N(0,\Sigma_{g})\label{eq:gmm}\\
\varepsilon_{g,i,j} & \sim N(0,\sigma_{\varepsilon,g}^{2}).\nonumber 
\end{align}
where the symbols have the same meaning as defined for the mixed model,
but are cluster-specific. The marginal mean of a GMM is thus given
by
\begin{equation}
\mathbb{E}(y_{i,j}|\boldsymbol{u}_{g,i})=\sum_{g=1}^{G}\pi_{g}\left[\boldsymbol{x}_{i,j}\boldsymbol{\beta}_{g}+\boldsymbol{\boldsymbol{z}}_{i,j}\boldsymbol{u}_{g,i}\right].\label{eq:gmm_mean}
\end{equation}

The model parameters are commonly estimated using Maximum Likelihood
Estimation (MLE) via the expectation-maximization (EM) algorithm.
Due to the large degrees of freedom, the iterative EM procedure is
unlikely to find the optimal solution, and instead tends to converge
towards suboptimal solutions. A better solution can be found by fitting
the model many times from random starting points and selecting the
best fit from these candidates. Alternatively, the solution of simpler
models is used as a starting point, e.g., using a GBTM \citep{jung2008introduction}.

Although GMM is suitable for ILD much like GBTM, it is considerably
slower to compute due to the number of model parameters growing drastically
with the number of clusters. Consider that each cluster has a new
set of parameters $\boldsymbol{\beta}_{g}$, $\Sigma_{g}$, and $\sigma_{\varepsilon,g}^{2}$,
in addition to the cluster-specific random variables $\boldsymbol{u}_{g,i}$
\citep{twisk2012classifying}. The model complexity is typically reduced
to speed-up estimation by assuming that certain parameters are identical
across the different clusters (e.g. the residuals and variances).
These challenges also inspired a different approach to performing
a covariate analysis. While these could be included into the GMM,
for larger datasets it is more practical to estimate an unconditional
GMM, followed by a multinomial logistic regression of the covariates
based on the subject cluster membership, referred to as a three-step
approach \citep{asparouhov2014auxiliary3}. We address the three-step
approach in a more general context in Section \ref{sec:guide}.

\paragraph*{Bayesian estimation}

In a Bayesian approach, the model parameters $\boldsymbol{\theta}$
of the GMM are treated as a random variable, whereas in MLE a point
estimate is obtained \citep{gelman2013bayesian}. The posterior distribution
of the model parameters given the data can be computed using Bayes'
rule
\begin{equation}
\mathrm{Pr}(\boldsymbol{\theta}|\boldsymbol{Y})=\frac{\mathrm{Pr}(\boldsymbol{Y}|\boldsymbol{\theta})\mathrm{Pr}(\boldsymbol{\theta})}{\mathrm{Pr}(\boldsymbol{Y})},
\end{equation}
where $\boldsymbol{Y}$ denotes the dataset, $\Pr\left(\boldsymbol{Y}|\boldsymbol{\theta}\right)$
denotes the likelihood of observing the data under the given model,
$\Pr\left(\boldsymbol{\theta}\right)$ denotes the prior information
about the model parameters, and $\mathrm{Pr}(\boldsymbol{Y})$ denotes
the evidence for the model. As $\mathrm{Pr}(\boldsymbol{Y})$ is constant
over $\boldsymbol{\theta}$, it suffices to consider $\Pr\left(\boldsymbol{\theta}|\boldsymbol{Y}\right)\propto\Pr\left(\boldsymbol{Y}|\boldsymbol{\theta}\right)\cdot\Pr\left(\boldsymbol{\theta}\right)$
for statistical inference. Bayesian inference is most beneficial when
informative priors can be provided, as the ability to incorporate
domain knowledge into the parameter estimation through priors improves
model estimation under low sample sizes \citep{gelman2013bayesian}.
However, specifying informative priors could be challenging in an
exploratory cluster analysis setting, especially when a large number
of clusters is sought out.

Compared to MLE, Bayesian inference allows for the estimation of more
complex models involving a large number of parameters, for which numerical
integration is infeasible \citep{ansari2000hierarchical}. In a comparison
between MLE and Bayesian estimation, \citet{depaoli2013mixture} found
that the Bayesian approach resulted in an improved recovery of the
model parameters. \citet{serang2015evaluation} demonstrated the improved
parameter recovery and smaller standard errors for estimating nonlinear
trajectories, applied to reading development trajectories of children,
although they noted an increase in convergence problems.

Due to the identical definition of the clusters in the mixture, the
cluster ordering (i.e., labels) can change freely during sampling,
referred to as the label switching problem. This presents a problem
when attempting to interpret the cluster-specific posterior distribution
samples. A possible solution to label switching is to add constraints
to the model to ensure identifiability, such as enforcing an ordering
on the cluster intercepts $\beta_{1,0}<\beta_{2,0}<...<\beta_{G,0}$
\citep{sperrin2010probabilistic}.

\paragraph*{Advanced applications}

Growth mixture modeling is a powerful and flexible statistical approach
for exploratory analyses, which is likely why the method has been
applied extensively by researchers throughout the past two decades.
Moreover, the model is applicable to ILD for the same reasons as GBTM.
An example of an ILD application is seen in the work of \citet{shiyko2012poisson},
who proposed an approach to applying a Poisson-GMM to ILD to investigate
patient's daily number of smoked cigarettes. Many researchers have
adapted GMM to meet their analysis needs. We highlight some of the
proposed extensions here.

\subparagraph*{Type of response}

The method can be applied to different types of data such as binary,
categorical, ordinal, count, and zero-inflated data, requiring different
distributions for the response \citep{muthen2009growth}. \citet{proust2009joint}
demonstrated a joint application of GMM in modeling multiple longitudinal
outcomes with time-to-event data. While the response distribution
can be determined from the data, the distribution of the random effects
is more difficult to establish, as wrongly modeling the within-cluster
heterogeneity simply results in additional clusters \citep{bauer2007observations}.
The assumption of normally distributed subgroups has been reconsidered
in recent years. Alternative distributions such as a skewed-normal
or skewed-t have been proposed in order to account for the skewness
and thickness of the tails of the random effects distribution, resulting
in a GMM which is more robust to non-normal groups and group outliers
\citep{lu2014bayesian,muthen2015growth,wei2017extending}. A disadvantage
of support for thicker tails is that it results in an even larger
overlap between clusters than is already the case for a mixture of
normals.

\subparagraph*{Trajectory representations}

Many researchers have explored different temporal shapes and structures.
\citet{grimm2010nonlinear} investigated nonlinear trajectories in
the reading development of children using specific functions. Nonlinear
trajectories have also been estimated using regularized polynomials
\citep{shedden2008regularized}, fractional polynomials \citep{ryoo2017nonlinear},
and splines \citep{marcoulides2018analyzing,ding2019development}.
Researchers have also accounted for sudden changes over time using
piecewise trajectory representations, referred to as a piecewise GMM
(PGMM) \citep{li2001piecewise}. PGMMs have also been proposed to
handle multiple change points, change points determined by the model
\citep{liu2018piecewise,ning2018class}, and subject-specific change
points \citep{lock2018detecting}. The intervals between change points
can also be regarded as a possible state change. In a multiphase or
sequential-process GMM \citep{kim2012investigating,reinecke2015stage},
the latent class membership is estimated per interval. State change
patterns can then be assessed using latent transition models \citep{collins2010latent}.

\subparagraph*{Missing data}

In most analyses, missing data is assumed to be missing at random.
However, it is not uncommon for the missing data mechanism to affect
the longitudinal data process, resulting in biased estimates if unaccounted
for \citep{little1995modeling}. This type of missingness is referred
to as non-ignorable. \citet{lu2011bayesian} applied a Bayesian approach
to modeling a GMM where the missing data mechanism was dependent on
the cluster and observed covariates. A common source of non-ignorable
missing data is the premature and permanent drop-out from observation.
A detailed overview of adaptations to model these mechanisms is provided
by \citet{enders2011missing} and \citet{muthen2011growth}. More
recent approaches to missing data include a shared-parameter model
\citep{gottfredson2014modeling,cetin2018comparison}.

\paragraph*{Software}

GMM is available through several modeling programs, e.g. \textsc{Mplus}
\citep{muthen2012mplus}, \textsc{Latent GOLD} \citep{vermunt2016technical},
and the R packages \textit{OpenMx} \citep{boker2011openmx}, \textit{lcmm}
\citep{proust2017estimation}, \textit{mixAK}\footnote{\url{https://CRAN.R-project.org/package=mixAK}}\citep{komarek2014capabilities},
\textit{flexmix} \citep{grun2008flexmix}, and \textit{mixtools} \citep{benaglia2009mixtools}.
GMM can be estimated using a Bayesian approach in several software
packages, including \textsc{OpenBUGS}\footnote{\url{http://openbugs.net}}
\citep{lunn2009bugs}, \textsc{Jags}\footnote{\url{http://mcmc-jags.sourceforge.net}}
\citep{depaoli2016just}, and \textsc{Stan}\footnote{\url{http://mc-stan.org}}
\citep{carpenter2017stan}, all of which have interfaces to R. In
R, specific Bayesian models are implemented, for example, in \textit{mixAK}
\citep{komarek2014capabilities}, and \textit{brms}\footnote{\url{https://CRAN.R-project.org/package=brms}}
\citep{burkner2017brms}.

\paragraph*{Case study}

\begin{figure}
\caption{\label{fig:gmm_case}GMM case study analysis.}
\subfloat[\label{fig:gmm_case_bic}BIC per solution (lower is better).]{\includegraphics{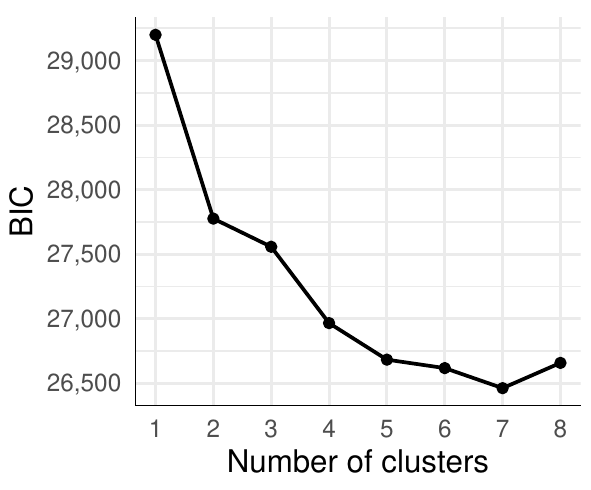}}\hfill{}\subfloat[\label{fig:gmm_case_groups}The identified cluster trajectories.]{\includegraphics{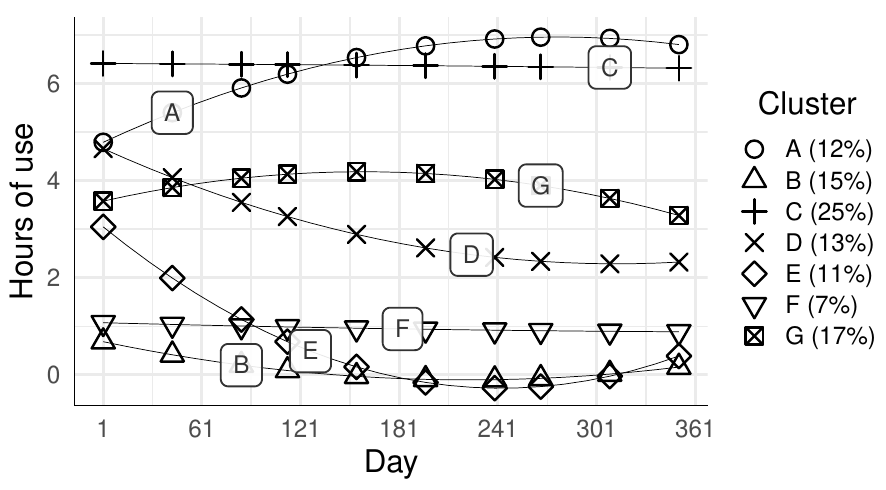}}
\end{figure}

The GMM analysis follows the same steps as the GBTM analysis, and
the same software is used to estimate the model. We therefore refer
to Table \ref{tab:lmm_case} for the exploration of the trajectory
shape using a single-cluster mixed model. We employ a quadratic GMM
with cluster-specific random patient intercepts, and cluster-independent
structured covariance matrices. We found that a grid search with 20
random starts was sufficient for consistently arriving at the best
solution. The resulting model BICs are shown in Figure \ref{fig:gmm_case_bic},
indicating that the best model fit is obtained at the seven-cluster
solution. The cluster trajectories thereof are shown in Figure \ref{fig:gmm_case_groups},
showing a close match with the ground truth insofar the cluster trajectories
can be represented using second-order polynomials.

\subsubsection{Time-varying effect mixture modeling}

In regression analysis, the associations between the covariates and
outcome are typically modeled as being constant over time. In practice
however, associations may change over time, resulting in a lack of
understanding of the true temporal association if this change is not
accounted for. In a varying-coefficient model (VCM), the dynamic association
between covariates is modeled using smooth functions \citep{hastie1993varying}.
VCM has been applied in longitudinal studies, in which covariates
are modeled with one or more time-varying coefficient functions denoted
by $\beta(\cdot)$. In this form, the model is referred to as a time-varying
coefficient model (TVCM), time-varying effect model (TVEM), or dynamic
generalized linear model. The individual trajectory is described by
\begin{equation}
y_{i,j}=\sum_{q=0}^{Q}x_{q,i,j}\beta_{q}(t_{i,j})+\varepsilon_{i,j},\label{eq:tvem}
\end{equation}
where $x_{0,i,j}=1$, $\beta_{0}$ denotes the time-varying intercept
over time, and $\beta_{q}$ denotes the temporal association between
the covariate $x_{q,i,j}$ and time. Furthermore, the residuals $\varepsilon_{i,j}$
are assumed to be normally and independently distributed with zero
mean and variance $\sigma^{2}$. The coefficient functions are described
through smooth continuous functions (i.e. the first-order derivative
is continuous), and are able to capture nonlinear longitudinal relations
between the covariates and time. Note that in the absence of covariates,
the model comprises a single coefficient function that captures the
longitudinal trajectory. TVEM is a promising approach for ILD, as
the large volume of data enables the identification of more complex
dynamic associations \citep{tan2012time}.

Over the years, several approaches have been suggested for the estimation
of the smoothing functions. Spline regression is used to describe
the function by a piecewise polynomial (typically of order 2 or 3)
over a given series of intervals \citep{liang2003relationship,hoover1998nonparametric}.
The interval boundaries, referred to as knots, need to be selected
carefully based on the data. An alternative approach named spline
smoothing does not require selection of intervals, but is much more
computationally intensive \citep{hoover1998nonparametric,hastie1993varying}.
A more recent approach involving P-splines takes the middle ground,
using a penalty factor to prevent overfitting while ensuring a smooth
fit \citep{song2010semiparametric,tan2012time}. Splines are described
through a linear model, and consequently, a TVEM describes a linear
model of which the model parameters can be estimated using ordinary
least-squares (OLS).

Mixtures of VCMs or TVEMs have been proposed for handling heterogeneity,
where the different groups are represented through clusters-specific
coefficient functions \citep{lu2012finite,dziak2015modeling,huang2018statistical,ye2019finite}.
\citet{lu2012finite} used an approach similar to GMM, where a random
intercept and slope are included in order to model within-cluster
heterogeneity. However, the use of linear random effects in combination
with nonlinear cluster trajectories may be limiting, as the nonlinear
changes remain homogeneous within cluster. \citet{dziak2015modeling}
proposed an alternative model which they named MixTVEM, given by
\begin{gather}
y_{i,j}=\sum_{q=0}^{Q}x_{q,i,j}\beta_{g,q}(t_{i,j})+\varepsilon_{g,i,j},\quad i\in I_{g}.\label{eq:mixtvem}
\end{gather}
The measured outcome $y_{i,j}$ is assumed to be normally distributed
when conditioned on the cluster variable. The model is similar to
GBTM, but accounts for cluster heterogeneity using an AR-1 model with
measurement error. \citet{huang2018statistical} proposed a mixture
of VCMs with flexible mixing proportions and dispersion, enabling
the modeling of these aspects over a covariate, e.g., time.

The parameters of the model can be estimated using the EM algorithm
\citep{dziak2015modeling} or a Bayesian approach \citep{lu2012finite}.
The optimization procedure for MixTVEM is initialized by assigning
random posterior probabilities to the classes. \citet{dziak2015modeling}
recommend to run the procedure for at least 50 random starts as the
optimization may converge on different solutions, or fail to converge
altogether. Due to the needed repeated runs, the tuning of the penalty
factor, and the relative complexity of the model, the method is highly
computationally intensive to estimate, as noted by \citet{yang2019performance}.

\paragraph*{Case study}

\begin{figure}
\caption{\label{fig:mixtvem_case}MixTVEM case study analysis.}
\subfloat[\label{fig:mixtvem_case_bic}BIC per solution (lower is better).]{\includegraphics{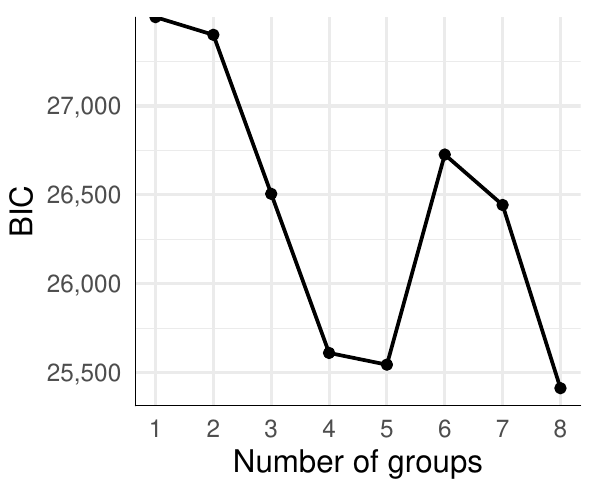}}\hfill{}\subfloat[\label{fig:mixtvem_case_groups}The identified cluster trajectories.]{\includegraphics{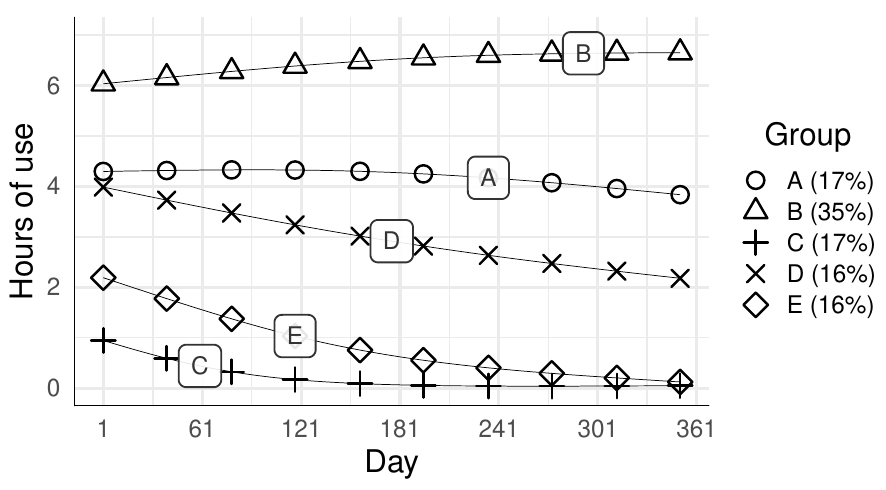}}
\end{figure}

The MixTVEM models are estimated using the R code provided by \citet{dziak2015modeling}\footnote{\url{https://github.com/dziakj1/MixTVEM}}
(version 1.2). P-splines of a third degree polynomial order are used
with six interior knots, spaced equally over time.. The model is fitted
from 20 random starts in order to obtain good starting conditions,
although on a few occasions, a rerun was needed due to convergence
problems. Moreover, the single-cluster estimation consistently failed
due to observations with zero variability from the non-users, which
we resolved by adding a negligible amount of perturbation to the measurements
with zero hours. The BICs of the selected solutions are depicted in
Figure \ref{fig:mixtvem_case_bic}, showing different levels of model
fit to the data across the number of clusters. We select the solution
involving five clusters as it best captures the different patterns
of change over time.

The five cluster trajectories are visualized in Figure \ref{fig:mixtvem_case_groups}.
Cluster B (35\%) comprises the good users and the slow improvers.
Cluster A (17\%) and D (16\%) correctly identify the variable users
and slow decliners, respectively, but the cluster trajectories do
not match the true shapes that would be expected from the correct
assignment. Cluster C (17\%) and E (16\%) comprise the non-users,
early drop-outs, and occasional attempters. However, the presence
of the occasional attempters appears to have affected both cluster
trajectories, such that neither matches the ground truth.

\subsection{\label{sec:numgroups}Number of clusters}

The determination of the number of clusters is a prominent topic
in the field of cluster analysis, as the number of clusters is usually
part of the model definition and can greatly affect the resulting
solution. However, there is no consensus on how to identify the true
number of clusters. This is largely attributable to the different
types of cluster analyses; each having different purposes, expectations,
and applications \citep{von2012clustering}. The identification of
the number of clusters is part of a broader search for the appropriate
cluster model, which we shall refer to as model selection. Interestingly,
considerable attention is given in the literature to the identification
of the number of clusters, over the more general topic of ensuring
the overall best model specification, referred to as model selection.
This is arguably justifiable in a longitudinal context under the assumption
that the trajectory models are sufficiently flexible. We summarize
the many approaches and metrics used for identifying the number of
clusters.

\paragraph*{Model metrics}

Although no tests exist for the number of clusters or the presence
of clusters, approximate likelihood ratio tests (LRT) enable researchers
to test whether the model with $G+1$ clusters describes the data
statistically significantly better than the identically specified
model with $G$ clusters. Commonly used variants are the Vuong-Lo-Mendell-Rubin
(VLMR) LRT \citep{lo2001testing}, the adjusted Lo-Mendell-Rubin (aLMR)
LRT \citep{lo2001testing}, and the bootstrap LRT (BLRT) \citep{mclachlan2000finite}.
These approaches are useful on smaller datasets for preventing overfitting.
However, the tests tend to result in the identification of too many
clusters (i.e., overextraction) on large datasets, where smaller changes
between models are statistically but not practically significant \citep{grimm2017model}.
A similar concept is seen in difference-like criteria, which measure
the relative improvement between successive cluster solutions \citep{vendramin2010relative}.
Along similar lines, \citet{grimm2017model} have applied $k$-fold
cross-validation for model selection based on how well the model represents
previously unseen data (in terms of the likelihood).

The most commonly applied approach involves the estimation of a cluster
model for a range of number of clusters. A metric is then used to
identify which of the models provides the best fit. Information criteria
strike a balance between model fit (the likelihood) and model complexity
(the number of model parameters). These model metrics can also be
used to compare across model specifications, selecting the model that
minimizes the metric.

Metrics for identifying the number of clusters have been studied extensively
for GMM and GBTM \citep{nylund2007deciding,feldman2009new,klijn2017introducing,tofighi2008identifying}.
Overall, the findings are mixed, likely due to the different settings
(e.g., sample size, cluster separation, noise) under which these evaluations
have been performed \citep{grimm2017model}. Overall, the BIC is commonly
used for the class enumeration in GBTM and GMM. The BLRT has been
demonstrated to be a reliable alternative \citep{nylund2007deciding,mcneish2017effect}.

\citet{malsiner2016model} proposed a metric based on the occurrence
of empty clusters in a mixture model with many clusters, where the
true number of clusters is determined based on the number of non-empty
clusters. This approach has the advantage of only requiring a single
model to be fitted. \citet{nasserinejad2017comparison} experimented
with this metric for different thresholds for the number of trajectories
that constitute a non-empty cluster. Another metric of interest is
the entropy of the posterior probability matrix, as a measure of cluster
separation (i.e., probabilities should be close to either zero or
one).

\paragraph*{Bayesian metrics}

Different criteria have been proposed for models estimated through
Bayesian inference. They make use of the posterior distribution of
the model coefficients. One of the more commonly used criteria is
the deviance information criterion (DIC), introduced by \citet{spiegelhalter2002bayesian}.
Its use has not been without criticism. For example, there exist multiple
definitions of the DIC, each with a different interpretation \citep{celeux2006deviance}.
\citet{spiegelhalter2014deviance} have summarized and addressed the
concerns. Recent alternative criteria are the widely applicable AIC
(WAIC), and Pareto-smoothed importance sampling using leave-one-out
cross validation (PSIS-LOO) \citep{vehtari2017practical}.

\paragraph*{Cluster criteria}

Criteria for cluster algorithms tend to assess the solution based
on the underlying data, and assume a hard partitioning of the data.
The advantage of such an approach is that it is independent from the
method that was used, making it possible to assess the model fit across
cluster methods. The criteria tend to contrast the within-cluster
variability against the between-cluster variability in order to assess
the separation between clusters, as seen, e.g., the Calinski-Harabasz
(CH) and Davies-Bouldin criteria. As an example, \citet{todo2016fitting}
found the CH criterion to perform better for model selection than
BIC in a latent profile analysis. Another commonly used criterion
is the ASW. A comprehensive overview of commonly used cluster criteria
is provided by \citet{vendramin2010relative}.

\paragraph*{Upper bound}

The upper bound on the largest number of clusters to be evaluated
is based on multiple factors. Firstly, prior knowledge may give researchers
reasons to expect the true number of clusters to be below a certain
number. Computational factors are also at play \citep{nasserinejad2017comparison},
as the model computation time scales non-linearly with an increasing
number of clusters for complex models, making the evaluation of a
larger number of clusters impractical. Along similar lines, the increasing
model complexity with an increasing number of clusters tends to result
in more frequent occurrences of convergence issues. The largest number
of clusters that can be estimated is also limited by the sample size,
considering that all submodels must comprise a sufficient number of
trajectories in order to obtain reliable estimates \citep{sterba2012factors}.
Studies involving a small sample size are therefore naturally limited
to identifying a lower number of clusters. Similarly, having a large
number of clusters limits the power of a post-hoc cluster comparison.

\paragraph*{Subjective assessment}

Researchers have argued against the optimization of a sole metric
for the identification of the number of clusters, as it is rather
mechanical in nature, and disregards the domain-dependent aspect of
the analysis \citep{nagin2018group}. Moreover, the commonly used
metrics tend to focus on discerning a sufficiently improved model
fit, however, a better model may fit aspects of the heterogeneity
which are not of interest for the purpose of the analysis \citep{van2019latent}.
This issue can occur in datasets with considerable overlap between
clusters, where the introduction of additional clusters may consistently
improve the model fit, albeit with diminishing returns. Due to the
non-linear nature of these diminishing improvements, there tends to
be a point or region where the marginal improvement drops. The identification
of this turning point, representing the preferred number of clusters,
creates room for subjectivity into the decision. This approach, often
assessed visually, is commonly referred to as the \textquotedbl elbow
method\textquotedbl . It is used, for example, by \citet{dziak2015modeling}
in their MixTVEM analysis, to assess the relative improvement in terms
of the BIC.

Arguably, the choice of metric or metrics involves a domain-dependent
decision. As the choice of the best metric may not be clear-cut, taking
into consideration multiple metrics can provide a more reliable result
\citep{ram2009methods}. However, because fit metrics capture different
aspects of the model fit, it is inevitable that some of the metrics
are in disagreement on the optimal number of clusters.

\paragraph*{Hierarchical models}

There is a practical limit to the number of clusters that can be used
to approximate the heterogeneity, for it becomes increasingly difficult
to produce unique labels for each of the clusters \citep{sterba2012factors}.
Instead of identifying an independent set of clusters, one can search
for a cluster tree hierarchy using a hierarchical cluster algorithm,
where each cluster is further explained in terms of subclusters. In
this way, an arbitrary level of granularity can be obtained up to
the subject level. This approach can be estimated through a cross-sectional
or feature-based approach using an agglomerative hierarchical cluster
algorithm. In recent years, \citet{van2017building} proposed a top-down
parametric approach based on GBTM, named latent-class growth trees
(LCGT). They identified the root of the hierarchy using a standard
GBTM analysis and metric, but subsequent clusters are fitted using
a two-cluster GBTM until no more significant improvement is obtained.
Another advantage of the approach is that the tree accounts for classification
error, as opposed to the hard partitioning used in traditional hierarchical
cluster algorithms. Furthermore, covariates can be included for comparison
or cluster membership prediction through a three-step estimation approach
\citep{van2019latent}.

\paragraph*{Non-parametric mixture models}

A promising alternative to the post-hoc identification of the number
of clusters, or model selection in general, is seen in the area of
nonparametric modeling, where only a single model is estimated. Here,
the model complexity is grown as needed to represent the data in an
infinite parameter space. In such a model, the number of clusters
$G$ is part of the model parameters to be estimated \citep{richardson1997bayesian,green2001modelling}.
This model can be realized using a Bayesian approach by placing a
Dirichlet process (DP) \citep{ferguson1973bayesian} prior on the
number of clusters $G$. The DP mixture model (DPMM) describes the
observations as a function of the model parameters $\boldsymbol{\theta}$
provided by the DP \citep{lo1984class}. It has been applied to clustering
gene expression data \citep{sun2017dirichlet}. \citet{heinzl2013clustering}
demonstrated that a DPMM could also be estimated with an EM algorithm
instead of MCMC, although they did not compare between estimation
algorithms. DPMMs can be estimated in R for example via the \textit{DPpackage}\footnote{\url{https://CRAN.R-project.org/package=DPpackage}}
by \citet{jara2011dppackage} or the \textit{BClustLonG}\footnote{\url{https://CRAN.R-project.org/package=BClustLonG}}
package by \citet{sun2017dirichlet}.

\section{Guidelines for conducting a longitudinal cluster analysis\label{sec:guide}}

Many decisions are involved in a longitudinal cluster analysis. The
need for guidelines comes not only from obtaining reliable results,
but also comes from ensuring proper reporting to enable reproducible
research. Unfortunately, the exploratory and domain-dependent nature
of clustering inevitably means that there is no single formal process
or method that covers all applications and purposes \citep{nagin2010groupclinical}.
Instead, the analysis should be adapted to the research questions
or intended application of the model. Guidelines can still play a
role here, as there are common themes to any longitudinal cluster
analysis.

We broadly outline the typical aspects and approaches involved in
a longitudinal cluster analysis. We focus on the guidance given by
researchers on the topic of mixture modeling, as these typically parametric
models tend to involve many decisions \citep{nagin2010groupclinical}.
We summarize the steps as follows:
\begin{enumerate}
\item Analyzing the model variables. The type of longitudinal response (e.g.,
categorical, ordinal, continuous) and the distribution thereof (e.g.,
normal, Poisson, zero-inflated, truncated) should be understood. In
addition, the distribution of the covariates should be investigated,
as outliers may skew the results.
\item Investigating the missing data mechanism. This step is crucial for
ILD, where the varying continuous measurement times may be underlying
to patterns of missingness. An advantage of clustering is that for
a missing data mechanism related to the longitudinal outcome, this
is handled by the clusters. Data is therefore usually assumed to be
missing at random (MAR). Missing not at random (MNAR) data has been
handled by using pattern mixture models.
\item Single-cluster modeling. Prior to the cluster analysis, it is good
practice to understand the performance of the single-cluster case
\citep{ram2009methods,van2017grolts}. If the single-cluster model
achieves a good fit, there may be little added value from complicating
the analysis by introducing additional clusters. Alternatively, the
heterogeneity could be assessed by comparing the coefficients obtained
from separate models for each trajectory if the sample size allows
for it. The single-cluster model may also be of use for identifying
the approximate trajectory shape.
\item Providing a rationale for clustering. Ideally, the analysis is justified
by theory or domain knowledge (e.g., previous studies) that strongly
hint at the existence of clusters. This step also pertains to the
way the clusters are interpreted, i.e., whether to use direct or indirect
clustering.
\item \label{enu:model}Identifying the best model. This step is by far
the most intricate, both in terms of number of decisions and computation
time. In view of exploring the data heterogeneity, it is preferable
to start with a model that does not account for covariates other than
time \citep{vermunt2010latent}, referred to as the unconditional
model. An example of method and model selection is found in the analysis
by \citet{feldman2009new}. The choice of method, model specification,
estimation method, and the selected number of clusters all affect
the model fit to the data. As such, arriving at the final model may
involve several iterations of the following substeps:
\begin{enumerate}
\item Choosing the cluster method. The methods have different strengths
and limitations in terms of, e.g., flexibility in modeling trajectory
shapes, capability to model heterogeneity, sample size requirements,
and computational scalability. It is worthwhile to weigh these aspects
in deciding on the method to use.
\item Choosing the estimation approach and method. Cluster models can be
challenging to estimate, as estimation algorithms may be unable to
identify the optimal solution in the vast parameter space. It is therefore
recommended to perform repeated runs with different random starting
values, and to select the model with the best fit from the candidate
models \citep{jung2008introduction,sher2011alcohol,mcneish2017effect}.
Moreover, it is worthwhile to experiment with different estimation
methods for improved convergence (e.g., by increasing the number of
iterations) and computational efficiency: The estimation algorithm
may fail to converge, or the identified solution is invalid due to
various reasons (e.g., out-of-bound coefficients, or empty clusters).
\item \label{enu:modelspec}Specifying and selecting the most appropriate
model. This typically manual process involves many decisions, including
the specification of the trajectory shape (e.g., polynomial, or spline),
the distribution of the response variable, any covariates, the shared
parameters between clusters (e.g., the covariance matrix), and cluster
heterogeneity. These decisions can be guided by domain knowledge or
by metrics for assessing the improved fit to the data. In particular,
the trajectory shapes can be explored using a cluster model with a
nonparametric representation of the cluster trajectory \citep{todo2016fitting}.
Alternatively, the task of model specification and selection can be
considerably simplified by using regularized or nonparametric models.
\item Identifying the number of clusters. There are many approaches to identify
the number of clusters, as described in \ref{sec:numgroups}. Typically
a forward selection approach is used where a cluster model is fit
and evaluated for an increasing number of clusters \citep{van2017grolts}.
One or more metrics, possibly taking into account domain knowledge,
are used to gauge the best number of clusters.
\item Assessing the model adequacy. The model fit can be assessed from the
residual observation errors of the model, which may reveal structural
deviations \citep{wang2005residual,feldman2009new,lennon2018framework}.
Adequacy may also be considered in terms of model parsimony, as similar
clusters or clusters representing only a small proportion of the trajectories
add little value to the overall model fit. The separation between
clusters can be evaluated through the cluster membership probability
matrix \citep{nagin2005group}, or by comparing the cluster trajectories
and the variability within clusters (either visually or by the means
or coefficients) \citep{feldman2009new,nagin2010groupclinical,lennon2018framework}.
It is also worthwhile to assess the standard errors or confidence
interval of the model coefficients for meaningful effects.
\item Validating the model. Longitudinal cluster models can involve many
parameters as the number of parameters scales linearly with the number
of classes, and thus the models are sensitive to overfitting (i.e.,
may not generalize well) on small datasets. If a model is estimated
on random subsets of the data (e.g., via bootstrapping) and yields
the same solution, this is indicative that the estimation of the model
is robust. Preferably, the model is evaluated on a holdout (i.e.,
validation) sample \citep{frankfurt2016using}, or using a $k$-fold
cross-validation approach. Here, the data is split into $k$ folds,
where $k-1$ folds are used for training, and the remaining fold is
used for testing. It is a useful approach for model evaluation or
selection under a more limited sample size \citep{grimm2017model}.
Overall, we observe few examples in literature where this step is
performed, nevertheless, it is advisable to assess the robustness
of the selected model, as an overspecification or overextraction of
the number of clusters may result in a model that does not generalize
well.
\end{enumerate}
\item Analyzing covariates. In many analyses, the association of the longitudinal
patterns with other variables is of interest. Covariates may be included
to explain the cluster membership or the variability within and between
clusters. There are different ways to go about analyzing these effects.
In a one-step approach, the covariates are included in the model specification
in step \ref{enu:modelspec}. The inclusion of covariates into the
model results in a more complex model which may be difficult to estimate,
leading to convergence issues or long estimation times. Moreover,
the interpretation of the identified longitudinal patterns becomes
more difficult, as the clusters are based on more than the longitudinal
change over time. In a standard three-step approach, the longitudinal
cluster model is first estimated without covariates to establish the
underlying latent groups. In the second step, individual trajectories
are assigned to a cluster. In the third step, the covariates are analyzed.
The last step can be approached in several ways. A post-hoc analysis
for comparing covariates between clusters is commonly done either
by comparing the means of covariates between clusters using ANOVA,
or by predicting cluster membership using multinomial logistic regression.
However, it is important to correct for the uncertainty in cluster
assignments when comparing covariates between clusters \citep{vermunt2010latent,bakk2013estimating}.
A more detailed overview of the different estimation approaches is
given by \citet{van2017grolts}.
\item Interpreting the findings. The implication of the identification of
clusters depends on the type of cluster application. A substantial
overlap between clusters may still yield meaningful findings in an
indirect application, yet discredit the existence of truly distinct
clusters for a direct application of clustering. Similarly, a predictive
application of the model with high accuracy depends on a large separation
between clusters. Most importantly, researchers should consider whether
the identified clusters or differences between clusters are statistically
and practically meaningful.
\end{enumerate}
With so many decisions involved in the analysis, reporting these decisions
is of the utmost importance. \citet{van2017grolts} developed a comprehensive
21-item checklist based on the consensus of 27 experts, referred to
as the guidelines for reporting on latent trajectory studies (GRoLTS),
with the aim of improving the transparency and replicability of the
analysis. Complementary to the guidelines summarized above, the checklist
recommends to report the software and version that was used to perform
the analysis, and to make the analysis source code available. While
we will not repeat the other items, we encourage the reader to read
the GRoLTS in full.

\citet{van2017grolts} conducted a preliminary analysis of the state
of reporting in the literature by applying GRoLTS to a selection of
studies. They selected 38 papers that used latent-class trajectory
modeling for identifying patterns of post-traumatic stress symptoms
after a traumatic event. On average, the papers only met nine of the
requirements, with the most complete paper meeting fifteen requirements.
We believe these findings help to quantify the broader problem across
domains of a lack of sufficient reporting. Guidelines such as GRoLTS
are therefore valuable and practical tools towards achieving greater
transparency, with more interpretable and reproducible findings.

\section{Discussion\label{sec:Discussion}}

The case study highlights the differences and similarities between
the evaluated approaches to longitudinal clustering. The most apparent
contrast is the different number of clusters of the best solutions
(either determined by a cluster metric or manual assessment). The
discrepancy is largely attributable to the different trajectory representations
and within- and between-cluster assumptions of the methods. All methods
converged on a solution for each of the requested number of clusters.
Moreover, the solutions for four clusters or less were highly similar
across methods.

The synthetic case study data comprised considerable between- and
within-patient variability. Despite this, the relatively straightforward
KML and LLPA approaches yielded useful solutions. While LLPA uses
the same non-parametric representational approach as KML, the identified
cluster trajectories were different. LLPA does account for variability
at each day allowing it to distinguish trend on the basis. In the
case study that resulted into detecting drop outs from attempters.
Both methods are fast to compute and involve a minimal number of modeling
decisions, and are therefore practical approaches for quickly obtaining
a sense of the variability in trajectory shapes in a heterogeneous
dataset. There are similarities to the solutions of KML and GBTM,
where KML is preferably for non-linear trajectories \citep{genolini2010kml}.
However, the ability to incorporate domain knowledge into a GBTM analysis
makes it suitable to assess heterogeneity even under small sample
size \citep{feldman2009new,twisk2012classifying}.

The solutions found by GBTM and MixTVEM were similar. However, MixTVEM
is more flexible yet conservative in the trend shapes due to its regularization,
which is generally preferable. The differences observed between the
GBTM and GMM solutions demonstrate the importance of the model specification.
Because of the large variation in intercept between patients, GBTM
needs more clusters to represent the many different patient trajectory
intercepts, whereas GMM can accommodate larger variability in intercepts
into a single cluster, leaving more clusters to model other temporal
differences (e.g., slope). However, this advantage comes at the cost
of a more complex model, resulting in significantly longer computation
times, and possibly convergence problems, as evident from the considerably
longer computation time of GMM over GBTM \citep{feldman2009new,twisk2012classifying}.
In a comparison between KML and GMM, it was found that GMM is preferred
\citep{twisk2012classifying}; this was the case even for a small
sample size \citep{martin2015growth}. Overall, the feature-based
approach and GMM most closely approximated the true group trajectories
from which the data was generated.

With the relatively recent attention for ILD, the number of studies
evaluating the methods on this type of data is limited however. This
is unfortunate as ILD presents new challenges with respect to the
volume of data, missing data, model complexity, and higher computational
demands. Many methods scale poorly with an increasing number of clusters,
placing practical limitations on the model complexity and volume of
the data. In case of large sample size or large number of observations,
this provides a serious practical limit on the maximum number of clusters
that can be estimated. An almost inevitable problem associated with
ILD is the missingness of data. Patterns for missingness. We only
briefly touched upon this topic.

Due to the broad scope of this tutorial, we cannot possibly cover
all areas of research on methods for longitudinal clustering. Nevertheless,
we do wish to mention some of these unaddressed areas. We restricted
the scope to a single outcome, whereas for example, KML, GBTM and
GMM have extensions that support multivariable longitudinal outcomes,
also referred to as joint trajectories. Furthermore, with the aim
of presenting the commonly used approaches to longitudinal clustering,
we may have omitted several alternative approaches. For example, we
only briefly touched upon the field of functional data analysis. This
is a class of methods that attempt to represent the data in terms
of smooth functions, a method to which TVEM is related. There has
been an increasing interest in further modeling sources of variation
in the data by modeling subject-specific variability in addition the
mean level, referred to as joint mean-variance modeling. As seen in
the LLPA case study demonstration, this can have an impact on the
identified clusters. In other applications, trajectories may be expected
to change cluster membership over time. Here, the clusters represent
different unobserved states in which the subject resides over time.
Here, a latent transition analysis can be used to model the transitions
between clusters \citep{collins2010latent}.

\section{Summary\label{sec:Summary}}

The area of longitudinal clustering has gained much traction over
the past two decades. We have attempted to present a comprehensive
guide on how longitudinal cluster analyses can be conducted, with
an emphasis on the different methods which are available for this
purpose. Clustering is a powerful tool for exploratory purposes, but
such analyses should be performed thoughtfully. We encourage researchers
to experiment with different methods and model specifications in order
to identify the most appropriate model for the data, and to report
the steps and decisions that were part of the analysis in order to
ensure interpretable results.

\section*{Acknowledgments}

The authors thank the anonymous reviewers for their valuable feedback
on this work.

\section*{Funding}

This work was supported by Philips Research. Niek Den Teuling and
Steffen Pauws are employees of Philips.

\section*{Supplementary materials}

The dataset and R code used in each of the examples is available online
at \href{https://github.com/philips-labs/demo-clustering-longitudinal-data}{https://github.com/philips-labs/demo-clustering-longitudinal-data}.

\bibliographystyle{chicago}
\bibliography{refs}

\appendix

\part*{Appendix}

\section{Strengths and limitations per approach}

\begin{table}[H]
\noindent \begin{centering}
\caption{High-level comparison between approaches.}
\par\end{centering}
\noindent \centering{}%
\begin{tabular}{>{\raggedleft}p{3cm}>{\raggedright}p{6cm}>{\raggedright}p{6cm}}
\toprule 
Approach & Strengths & Limitations\tabularnewline
\midrule
\midrule 
Cross-sectional clustering & \begin{itemize}
\begin{singlespace}
\item Fast to compute
\item Algorithm implementations are widely available
\item Non-parametric cluster trajectory representation
\end{singlespace}
\end{itemize}
 & \begin{itemize}
\item Observation moments must be aligned across trajectories
\item Requires complete data
\item Sensitive to measurement noise \citep{green2014latent}
\end{itemize}
\tabularnewline
\midrule 
Distance-based clustering & \begin{itemize}
\item Versatile; many available distance metrics, which could also be combined
\item The distance matrix only needs to be computed once
\item Fast to evaluate for a large number of clusters
\end{itemize}
 & \begin{itemize}
\item Only practical up to a limited number of trajectories, as the number
of pairwise distances to compute grows quadratically with the number
of trajectories
\item No robust cluster trajectory representation (centroid trajectory may
not be insightful)
\item Some distance metrics require aligned observations (e.g., Euclidean)
\end{itemize}
\tabularnewline
\midrule 
Feature-based clustering & \begin{itemize}
\item Versatile; longitudinal features can be arbitrarily combined into
a trajectory model
\item Fast to compute
\item Can incorporate domain knowledge
\item Compact trajectory representation
\end{itemize}
 & \begin{itemize}
\item Generally requires ILD in order to ensure a reliable estimation of
the features per trajectory
\item Feature estimates may be unreliable for trajectories that cannot be
represented
\end{itemize}
\tabularnewline
\midrule 
Mixture modeling & \begin{itemize}
\item Parametric cluster trajectory representation
\item Versatile; choice of latent-class model, trajectory model, latent-class
membership model
\item Compact trajectory representation
\item Relatively low sample size requirement, both in number of trajectories,
and number of observations per trajectory \citep{martin2015growth}
\item Domain knowledge can be incorporated
\item Can assess the association of external variables or distant outcomes
\end{itemize}
 & \begin{itemize}
\item Computationally intensive
\item Number of parameters typically scales linearly with the number of
clusters
\item The estimation procedure may not converge to a good solution; many
random starts are needed
\end{itemize}
\tabularnewline
\bottomrule
\end{tabular}
\end{table}

\section{Strengths and limitations of mixture models}

\begin{table}[H]
\noindent \begin{centering}
\caption{High-level comparison between the described mixture models.}
\par\end{centering}
\noindent \centering{}%
\begin{tabular}{>{\raggedleft}p{3cm}>{\raggedright}p{6cm}>{\raggedright}p{6cm}}
\toprule 
Approach & Relative strengths & Relative limitations\tabularnewline
\midrule
\midrule 
GBTM & \begin{itemize}
\begin{singlespace}
\item Fast to compute
\end{singlespace}
\item Few parameters
\item Easy to interpret
\end{itemize}
 & \begin{itemize}
\item Sensitive to outliers
\item Poor fit, as individual trajectories are not modeled
\item Tends to overestimate the number of clusters \citep{twisk2012classifying}
\end{itemize}
\tabularnewline
\midrule 
GMM & \begin{itemize}
\item Within-cluster heterogeneity
\item Fewer clusters needed to represent heterogeneity \citep{muthen2015growth}
\item Random effects allow for cluster trajectories with a lower emphasis
on, e.g., intercept.
\item Forecast individual trajectories
\end{itemize}
 & \begin{itemize}
\item Slow to compute \citep{twisk2012classifying}
\item Requires many random starts \citep{mcneish2017effect}
\item Convergence issues \citep{twisk2012classifying}
\item Clusters can overlap considerably \citep{feldman2009new}
\item Sensitive to the specified distribution of the random effects 
\end{itemize}
\tabularnewline
\midrule 
MixTVEM & \begin{itemize}
\item Easy to interpret
\item Assess time-dependent association of external variables
\item Penalized splines result in less spurious temporal patterns
\end{itemize}
 & \begin{itemize}
\item Slow to compute \citep{yang2019performance}
\item Requires tuning of penalization factor
\item Convergence issues \citep{yang2019performance}
\end{itemize}
\tabularnewline
\bottomrule
\end{tabular}
\end{table}

\end{document}